\documentclass[sigconf]{acmart}

\usepackage[normalem]{ulem}
\usepackage{amsmath,amsfonts}
    \usepackage{algorithmic}
\usepackage{graphicx}
\usepackage{subfig}
\usepackage{textcomp}
\def\BibTeX{{\rm B\kern-.05em{\sc i\kern-.025em b}\kern-.08em
    T\kern-.1667em\lower.7ex\hbox{E}\kern-.125emX}}

\newtheorem{goal}{\textbf{Design Goal}}

\usepackage{soul} 
\usepackage{amsfonts}

\usepackage{diagbox}
\usepackage[linesnumbered,ruled,vlined]{algorithm2e}
\definecolor{verde}{rgb}{0.25,0.5,0.35}
\definecolor{jpurple}{rgb}{0.5,0,0.35}
\definecolor{darkgreen}{rgb}{0.0, 0.2, 0.13}
\usepackage{listings}

\definecolor{almond}{rgb}{0.98, 0.98, 0.82}
\definecolor{blue}{rgb}{0.0, 0.0, 1.0}

\usepackage{tikz}

\usepackage{enumitem}
\newenvironment{myitemize}
{ \begin{itemize}[leftmargin=0.2in]	
		\vspace{-1ex}	
		\setlength{\itemsep}{0pt}
		\setlength{\parskip}{0pt}
		\setlength{\parsep}{0pt}    }
	{ 	 \end{itemize}                    }

\newenvironment{myenumerate}
{ \begin{enumerate}[leftmargin=0.2in]
		\vspace{-1ex}
		\setlength{\itemsep}{0pt}
		\setlength{\parskip}{0pt}
		\setlength{\parsep}{0pt}    }
	{ \end{enumerate}                  }

\newcommand{\system}{\texttt{PLStream}\xspace}

\usepackage{hyphenat}
\newcommand{\compact}{\vspace{-5pt}}
\newcommand{\subcompact}{\vspace{-5pt}}

\usepackage[skip=3pt]{caption}
\begin{document}
\title[A Framework for Fast Polarity Labelling of Massive Data Streams]{A Framework for Fast Polarity Labelling of Massive Data Streams}

\author{Huilin Wu}
\author{Mian Lu}
\author{Zhao Zheng}
\affiliation{%
  \institution{4Paradigm Inc.}
  \country{}
}
\email{[wuhuilin,lumian,zhengzhao]@4paradigm.com}



\author{Shuhao Zhang}
\affiliation{%
  \institution{Singapore University of Technology and Design}
   \country{}
}
\email{shuhao\_zhang@sutd.edu.sg}


\begin{abstract}
Many of the existing sentiment analysis techniques are based on supervised learning, and they demand the availability of valuable training datasets to train their models.  
When dataset freshness is critical, the annotating of high speed unlabelled data streams becomes critical but remains an open problem.
In this paper, we propose \system, a novel Apache Flink-based framework for fast polarity labelling of massive data streams, like Twitter tweets or online product reviews.
We address the associated implementation challenges and propose a list of techniques including both algorithmic improvements and system optimizations.
A thorough empirical validation with two real-world workloads demonstrates that \system is able to generate high quality labels (almost $80\%$ accuracy) in the presence of high-speed continuous unlabelled data streams (almost $16,000$ tuples/sec) without any manual efforts.
\end{abstract}

\ccsdesc[500]{Computing methodologies~Massively parallel algorithms}
\ccsdesc[500]{Computing methodologies~Natural language processing}

\keywords{sentiment classification, stream processing, distributed computing}

\maketitle

\section{Introduction}
\label{sec:intro}
Companies have utilized sentiment analysis~\cite{capuano2021sentiment} in many business services to gain useful insights, such as the opinions of new products.
Although unsupervised methods exist, many sentiment analysis tasks still work with supervised learning models~\cite{56622554082144859b6c5cf5ff238ad8}. 
Their effectiveness thus highly depends on the quantity and quality of training data.
Despite the popularity of applying unsupervised pre-trained models such as GPT-3 and BERT~\cite{devlin2018bert} in sentiment analysis, a supervised downstream fine-tuning task is still required to achieve outstanding performance, such as model fine-tuning for sentiment analysis. 
A large amount of labelled datasets is still required, and it is costly to label the datasets with human efforts.


Motivated by such a demand, Iosifidis et al.~\cite{iosifidis2017large} have recently introduced a semi-supervised labelling approach to annotate polarity labels. However, their are focusing on theoretical accuracy guarantees with self-learning and co-training techniques, but overlook the annotating speed as well as the generality of the algorithm. 
In particular, it is a well-established industry practice to use the most recent data to train machine learning models~\cite{10.1145/3447548.3467172}. 
New datasets are continuously generating in the form of data streams, such as twitter tweets and online customer reviews. 
When dataset freshness is critical to sentiment analysis, frequent offline label retraining and deployment are constrained by the cost of manual labelling, limited computation and storage resources. 

\begin{figure}[t]
    \centering
    \includegraphics[width=0.4\textwidth]{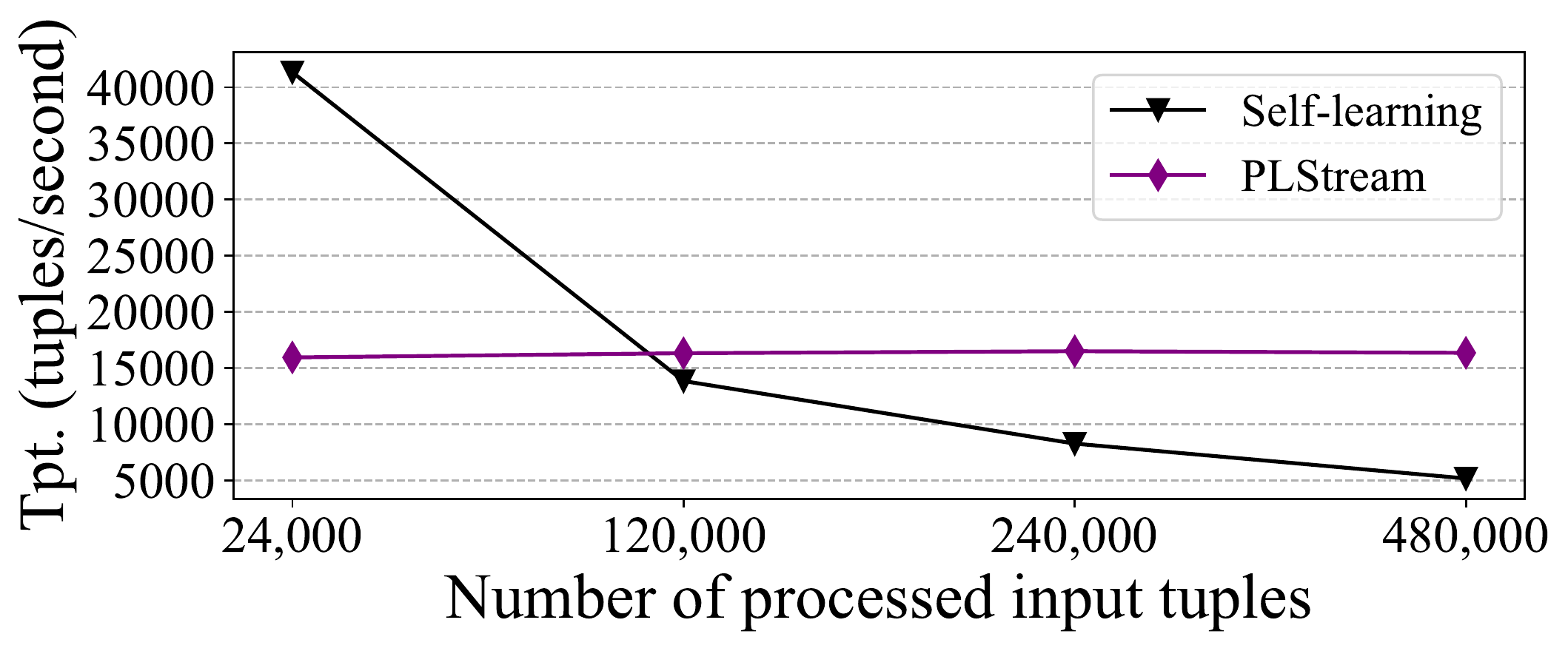}
    \caption{Comparing the performance of polarity labelling of increasing volume of datasets on 48 cores: an existing approach (Self-learning), and our framework \system.}
    \label{fig:selflearn}
\end{figure}

Figure~\ref{fig:selflearn} further shows the evaluation results of the existing self-learning approach~\cite{iosifidis2017large} to annotate polarity labels of tweets.
We let it first train an initial model with 5000 labelled data.
Then, it
uses the high-confidence predicted results as new training data for iterative supervised training of the model, where the rest unlabelled data (with less confidence) and new input data are recollected for the next training iteration. 
We can see that, the existing self-learning approach is incapable in handling data streams as it involves more and more datasets for training leading to an exponentially growing training time. 
It is not able to make any further progress when number of processed input tuples exceed few hundred thousands of tuples (i.e., tweets), but input data stream is potentially infinite.
Henceforth, we do not further evaluate such self-learning approaches, and a new approach to continuously annotate high volumes of data streams is required.

In this work, we propose \system, a novel framework that
\emph{continuously annotates high quality polarity labels to massive ongoing streams of opinions}. 
As Figure~\ref{fig:selflearn} shows, \system is able to achieve steady high throughput when annotating polarity labels of data streams. The key of \system is a novel online sentiment learning paradigm, which updates a word-embedding model (e.g., word2vec~\cite{mikolov2013efficient}) incrementally and generates labels by referencing the learned vector representations to a small reference table of vocabularies with assumed polarities. A list of algorithmic improvements and system optimization are further developed to help \system achieve both high performance and high labelling correctness. 
To ensure its high scalability and robustness, we have implemented \system on top of Apache Flink~\cite{carbone2015apache} -- a production level data stream processing system (DSPS). Nevertheless, our proposed solution is general and can be applied to other DSPSs such as Spark~\cite{zaharia2012resilient} with marginal programming efforts. 

For a comprehensive comparison, we have further implemented several alternative methods, including
statistics-based~\cite{lin2009joint}, lexicon-based~\cite{ohana2009sentiment} and clustering-based~\cite{li2014sentiment} unsupervised sentiment classification algorithms in \system for labelling data streams without labelled datasets.
The experimental evaluation shows that our proposed solution is able to continuously generate high quality labels (almost $80\%$ accuracy) to high-speed continuous data streams (almost $16,000$ tweets/s).
Compared to the alternative approaches, \system improves labelling accuracy by more than $20\%$, and achieves more than $50\%$ improvements in terms of both latency and throughput.

In summary, we make the following technical contributions:
\begin{myenumerate}
    \item
        We design a scalable and robust framework (Section~\ref{sec:overview}), \system for annotating unlabelled data streams based on a production-level data stream processing system, Apache Flink. 
        We open-source our system\footnote{\url{https://github.com/HuilinWu2/PLStream}}.
    \item 
        We propose and evaluate several algorithmic and system optimizations in \system including 
        \emph{(i) fast similarity-based labelling} (Section~\ref{subsec:labelling}),
        \emph{(ii) temporal trend detection} (Section~\ref{subsec:detection}), and
        \emph{(iii) tunable model management} ((Section~\ref{subsec:model_management}))
         that help \system to achieve high performance and high labelling accuracy.
    \item
        To the best of our knowledge, this is also the first empirical study that ensures a fair comparison among alternative approaches on annotating unlabelled data streams on modern parallel hardwares (Section~\ref{sec:experiment}), as we eliminate the performance difference caused by different compilers and runtime by implementing all alternative solutions inside the same framework. 
\end{myenumerate}

\compact
\section{Related Work}
\label{sec:related}
We discuss related work comes from three key areas: sentiment classification, large scale annotations and stream processing.

Sentiment classification is an application of natural language processing (NLP) that analyzes subjective texts with emotions to determine the views, preferences, and tendencies of a text.
Most prior works~\cite{2016Machine,go2009twitter,gautam2014sentiment,haque2018sentiment,ortigosa2014sentiment} belong to \emph{supervised sentiment learning}, which typically requires a \emph{training} phase that trains a classifier with labelled training texts followed by an \emph{inference} phase that identifies the polarity of a testing text based on the trained classifier. Their primary concern is to obtain a high accuracy classifier from a large set of collected and labelled datasets, assuming infinite time and resource budgets.
Zimbra et al.~\cite{56622554082144859b6c5cf5ff238ad8} reviewed 28 existing techniques of sentiment classification across five distinctive twitter data sets, and are mostly supervised approaches. 
Unsupervised methods, such as lexicon-based approaches~\cite{2011Lexicon,baccianella2010sentiwordnet} have been proposed. However, lexicon-based approaches only deal with novel vocabulary and contextual information in a limited way as they rely on domain-dependent opinion words and aspect-dependent opinion words~\cite{6994276}. Probabilisitc models, such as LDA~\cite{lin2009joint}, target learning from a large set of words from documents, while we target processing data streams, like tweets, which are written in a unique conversational style particular to the brevity of the Twitter medium (i.e., 280 character limit per tweet). 

Large scale annotations of unlabelled datasets is not a new demand. 
Early work by Pak and Paroubek~\cite{pak-paroubek-2010-twitter} utilizes emoticons to infer labels from tweets and is limited with their application scenarios.
Closely related to our work, a recent work~\cite{iosifidis2017large} applies Multinomia Naive Bayes (MNBs) as the based model to annotate sentiment datasets. 
Combined the MNBs model with a confidence threshold, they use the classification results as new training data to incrementally train the model offline until the accuracy converges.
However, we haven shown in Section~\ref{sec:intro} that, such an offline approach is not applicable to annotate high-speed data streams due to the exponentially increasing computational complexity.
To the best of our knowledge, how to ensure high accuracy of polarity labelling of massive unlabelled data streams such as social media opinions, so far has been an unsolved problem. %

Most existing sentiment analysis techniques have emphasized algorithmic accuracy. They have not focused on utilizing parallel hardware resources. 
Although there are general-purpose data stream processing systems (DSPSs) that can provide fast stream computations (e.g., Apache Spark streaming~\cite{Sparkstream} and Apache Flink~\cite{carbone2015apache}), they have not been designed nor optimized for the tasks of sentiment analysis. 
This had motivated a distributed online sentiment analysis system as proposed by Amir et al.~\cite{rahnama2014distributed}. Liu et al.~\cite{Liu2015FLORINA} have described a standalone system that supports real-time unsupervised machine learning tasks such as lexicon-based sentiment analysis. We evaluate the lexicon-based sentiment classification as an alternative approach in our experiments.
Recently, Ramanath et al.~\cite{10.1145/3447548.3467172} have proposed a library called Lambda Learner, which trains sequential bayesian logistic regression models by incremental updates in response to mini-batches from data streams. In contrast, our work strictly focuses on fast polarity labelling of massive data streams.


\section{Preliminary}
\label{sec:background}
In this section, we present the preliminary knowledge of basic concepts in word embedding and the data stream processing systems.

\textbf{Word Embedding.}
To analyze unstructured natural language texts, a common step is to first convert vocabulary into a vector of numeric values.
Two classical word representation algorithms are \emph{one-hot encoding} and \emph{bag of words}~\cite{birunda2021review}. However, one-hot encoding will lead to a dimension explosion when the vocabulary size becomes larger, while bag of words does not consider the context information in one sentence. 
Another advanced word vector representation technique, \emph{word embedding}, is to find a mapping or function to generate an expression in vector space. Representative recent word embedding tool include \emph{Word2Vec}~\cite{mikolov2013efficient},  \emph{Glove}~\cite{2014Glove}, \emph{BERT}~\cite{devlin2018bert} and their variants~\cite{10.5555/3044805.3045025}. 
\emph{Glove} combines local contextual information with overall statistics of the corpus, while word embedding with pre-trained model \emph{BERT} provide richer embedded vectors, where \emph{BERT} uses deep bidirectional transformer~\cite{vaswani2017attention} components to build the entire model with bidirectional contextual information, i.e., each word has different vector representations in different context. However, the training of \emph{Glove} and \emph{BERT} are batch-based, requiring the presence of the entire input datasets and cannot be applied to learn from on-going input streams.
In contrast, Word2Vec~\cite{mikolov2013efficient} is able to learn incrementally, and is thus adopted in our framework.

\textbf{DSPSs.}
With the explosion of information, being able to process data streams efficiently in real time is perceived as a major challenge. 
In the last decade, data stream processing systems (DSPSs) have received considerable attention. Many DSPSs have been proposed, such as Apache Flink~\cite{carbone2015apache}, Apache Storm~\cite{ApacheStorm}, and Spark Streaming~\cite{Sparkstream}.
We leverage Apache Flink for stream processing due to its capability of expressive, declarative, and efficient data analysis on data streams~\cite{carbone2015apache}, with high-level programming language API support such as the Python API~\cite{pyflink}. 
Nevertheless, our proposed solution is general and can be applied to other DSPSs with similar characteristics. %
\compact
\section{Incremental Polarity Learning}
\label{sec:overview}
We propose \system, a novel framework that learns and identifies the polarity of fast ongoing data streams.
We summarize the notations used throughout this paper in Table~\ref{tab:notations}.
We define a \emph{tuple} (denoted as $T$) as a timestamp $t$ associated with a finite number of vocabularies (i.e., $v_1\sim v_n$). The \emph{input stream} is a list of tuples chronologically arriving at the system, denoted as $S=\{T_1, ..., T_N\}$, where $N$ goes to infinity. In \system, every input tuple functions as both the \emph{training} and \emph{testing} datasets. 
The goal of \system can be thus defined as follows.
\begin{goal}
Given an input stream $S$, \system learns and identifies the polarity of $T_i \in S$ as soon as $T_i$ arrives without relying on any offline training phase with labelled datasets. 
\end{goal}

\begin{figure}[t]
    \centering
    \includegraphics[width=0.47\textwidth]{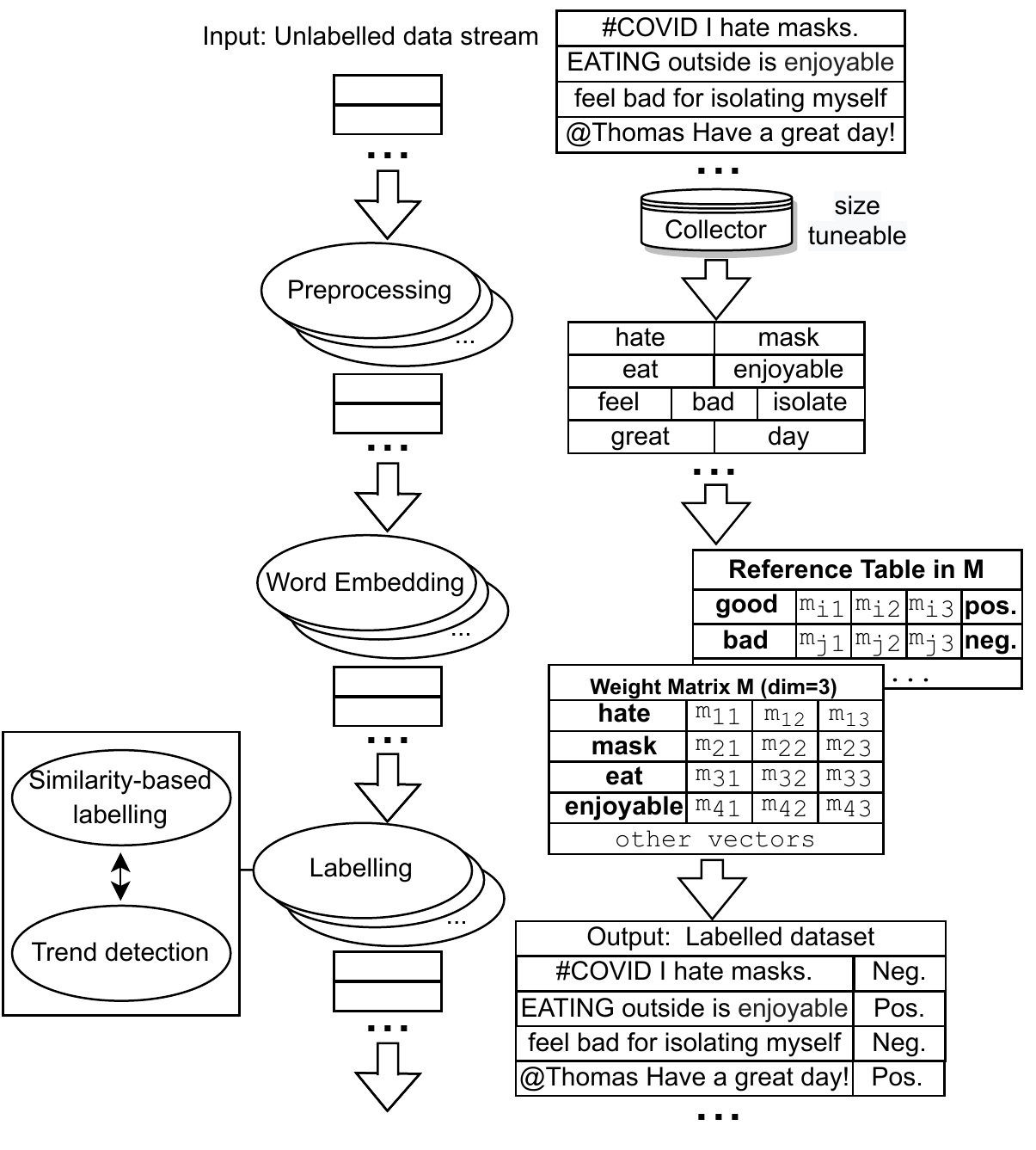}
    \caption{System overview. An oval denotes one instance of a stream operator.}
    \label{fig:system_overview}
\end{figure}

An overview of \system is depicted in Figure~\ref{fig:system_overview}. 
It is implemented on top of Apache Flink as a chain-topology stream application with three stream operators. Each operator can be carried by multiple threads to run concurrently to improve system throughput.
It is noteworthy that, \system is continuously updated and can be used to label sentences with even unforeseen vocabularies. This differs significantly from offline approaches~\cite{2016Machine,haque2018sentiment}, where the pre-trained model can quickly become outdated, as both the vocabulary and the polarity model evolve, requiring periodically re-training with labelled datasets.

First, it continuously batches a tunable number ($b$) of tuples from the input data stream and extracts useful vocabulary to feed to the subsequent operators.
To reduce the processing workload, information that have nothing to do with emotions, such as numbers, stopwords, usernames, etc. is filtered out. 
For example, in Figure~\ref{fig:system_overview}, $\#COVID$, $I$ and $!$ are removed because they are unrelated to sentiment.
It is noteworthy that we retain as much as possible of the words that constitute the sentence to provide contextual information. This contradicts to some prior work~\cite{2011Lexicon}, which only retain adjectives and adverbs. For example, the noun ``criminal'' is more likely to appear in negative texts.

Second, \system performs an incremental word embedding based on Word2Vec as it captures the vector representation of each vocabulary with local information involving low computational costs compared to alternative embedding approaches like BERT~\cite{devlin2018bert}. 
It continuously trains and updates vector representations (VRs) of input tuples and help make high-accuracy polarity predictions by extracting both \emph{contextual and temporal information} from opinionated streams. For example, in Figure~\ref{fig:system_overview}, weight matrix $M$ is incrementally trained and each row stands for the VR for one word.

Third, with the learned vector representation of each word, \system classifies the polarity of each input tuple without manual labelling efforts. For example, in Figure~\ref{fig:system_overview}, the polarity of four input tuples (i.e., tweets) are labelled through similarity-based computations. To improve the system performance and labelling accuracy, \system adopts a fast cosine similarity computation method to estimate the polarity label of the sentences and enhance it further with a temporal trend detection mechanism, to be covered in Section~\ref{subsec:labelling} and~\ref{subsec:detection}, respectively. 

During the second step, 
the learned VRs may be stored locally by each thread or in Redis for shared access by multiple threads.
When VRs are shared among threads, continuously updating VRs potentially leads to huge overhead, especially the communication overhead between two systems, i.e., Flink and Redis.
To this end, we further develop several alternative model merging approaches, detailed in Section~\ref{subsec:model_management}.

 

\begin{table}[]
\centering
\caption{Notations used in this paper. Parameter value ranges used in our experiments are also shown.}
\label{tab:notations}
\resizebox{0.45\textwidth}{!}{%
\begin{tabular}{|p{1.5cm}|p{2.8cm}|p{4.5cm}|}
\hline
Name         & Notation & Description                                                                     \\ \hline
Tuple        & $T=\{t, v_1, v_2, ..., v_n\}$        & A timestamped ($t$) finite number $n$ of words ($v_i$) with a polarity to be identified \\ \hline
Input stream & $S=\{T_1, ..., T_N\}$     & A list of tuples chronologically arriving at the system to be analyzed \\ \hline
Weight Matrix & $M_{m,d}$, where each element in $M$ is denoted as $m_{i,j}$  & The vector representation of a list of words, where $m$=number of words, and $d$=number of features                                                                \\ \hline
Vector lookup table        & $V: word \rightarrow m_{i}$    & A lookup table of trained word (key) with corresponding vector representation in $M$ (value)                                                                         \\ \hline
Window        & $w=1\sim10$     & A tunable training window size used by Word2Vec to capture contextual information                                                                              \\ \hline
Vector Dimension        & $d=10\sim500$     & A tunable feature size used by Word2vec training                                                                    \\ \hline
Batch size        & $b=200\sim5000$     & A tunable size of batched tuples being collected before training                                                                    \\ \hline
Trend Detection Window        & $TDW=200\sim2000$     & A tunable window size for polarity trend detection                                                                               \\ \hline
Model Merging Periodic       & $p=20\sim90 (seconds)$     & Model merging frequency                                                                              \\ \hline
\end{tabular}%
}
\end{table}

\compact
\begin{table}[]
\caption{Reference table used in our experiments}
\begin{tabular}{|l|l|}
\hline
Positive & \begin{tabular}[c]{@{}l@{}}dazzling brilliant phenomenal excellent fantastic \\ gripping mesmerizing riveting spectacular cool \\ awesome thrilling moving exciting love wonderful \\ best great superb still beautiful\end{tabular} \\ \hline
Negative & \begin{tabular}[c]{@{}l@{}}sucks terrible awful unwatchable hideous bad \\ cliche boring stupid slow worst waste unexciting \\ rubbish tedious unbearable pointless cheesy \\ frustrated awkward disappointing\end{tabular}            \\ \hline
\end{tabular}
\label{tbl:reference}
\end{table}

\begin{figure}
    \centering
    \includegraphics[width=0.37\textwidth]{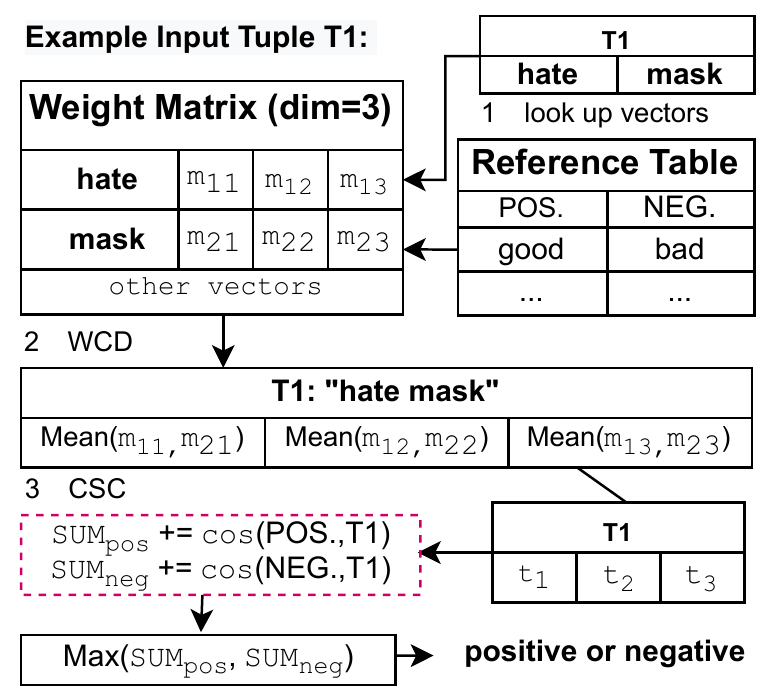}
    \caption{Example of similarity-based labelling}
    \label{fig:similarity}
\end{figure}
\subsection{Fast Similarity-based Labelling}
\label{subsec:labelling}
After getting the learned vector representation of each word, 
instead of relying on supervised training progress for polarity identification, we propose to rely on text similarity~\cite{text_similarity_1,text_similarity_2,text_similarity_3} to infer the polarity of input tuples based on a list of reference words with assumed polarity~\cite{lin2009joint} listed in Table~\ref{tbl:reference}. 
The reference table may become outdated as well. In practice, one may periodically update the table, which involves minor labor efforts compared to manually labelling training datasets.

Intuitively, we can compute the cosine similarity between each word of the input tuple and the reference words as the quantitative relative polarity of each word.
We can then compared the sum of the quantitative polarity of all words to get the polarity score of an input tuple.
However, such a simple solution leads to significant computational overhead as it involves multiple cosine similarity computations for every word in each input tuple. 

We leverage a simple yet effective \emph{word centroid distance (WCD)} calculation approach. 
For illustration, we show our running example in Figure~\ref{fig:similarity}. 
First, we obtain the vector representation of the input tuple by calculating the arithmetic mean of the learned vector representation of all words. 
As illustrated in the example, for the input tuple ``hate mask'', we calculate the mean of each dimension ($d=3$) of two words ($S=2$) to obtain a new vector $V= [v_1, v_2, ..., v_d], where\ v_i=\frac{\sum_{j=1}^{S}{m_i,j}}{S}, and\ d=3,\ S=2$.
Then, we perform cosine similarity computation (CSC) between $V$ and all reference positive or negative words to obtain the aggregated sum as $SUM_{pos}$ and $SUM_{neg}$. 
Finally, we compare $SUM_{pos}$ and $SUM_{neg}$ to infer the polarity of the input tuple entirely.
Our extensive experimental results confirm the superiority of such a simple solution in terms of both significantly reduced computational effort and higher prediction accuracy compared to alternative solutions. 

\begin{figure}[t]
    \centering
    \includegraphics[width=0.45\textwidth]{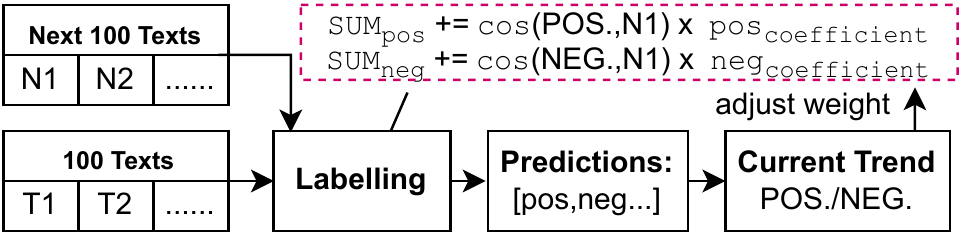}
    \caption{Example of trend detection}
    \label{fig:trend}
\end{figure}
\begin{figure*}[!ht]
\subfloat[Local model]{%
    \includegraphics*[width=0.27\textwidth]{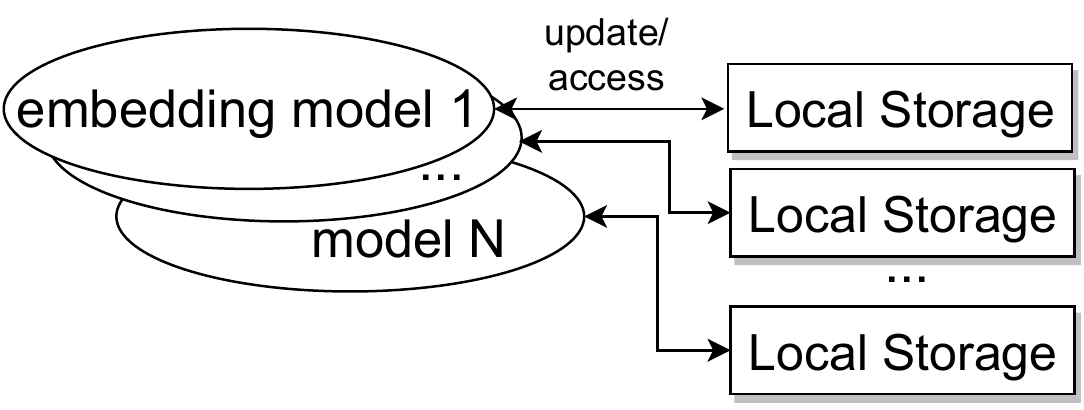}   
}
\hfill
\subfloat[Global model]{%
    \includegraphics*[width=0.27\textwidth]{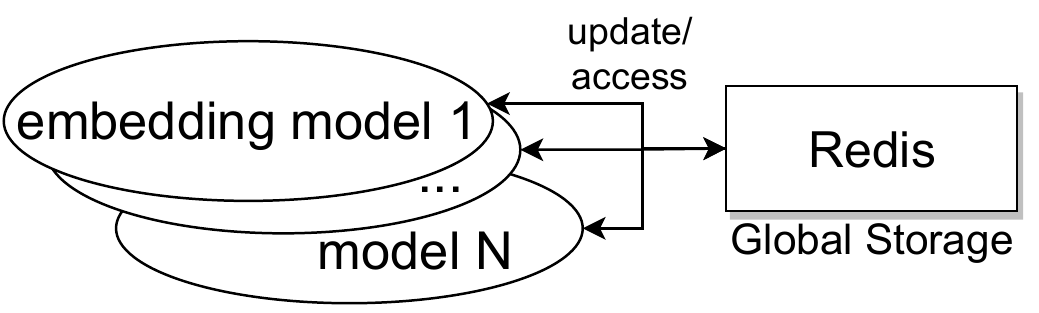}   
}
\hfill
\subfloat[Hybrid model]{%
    \includegraphics*[width=0.3\textwidth]{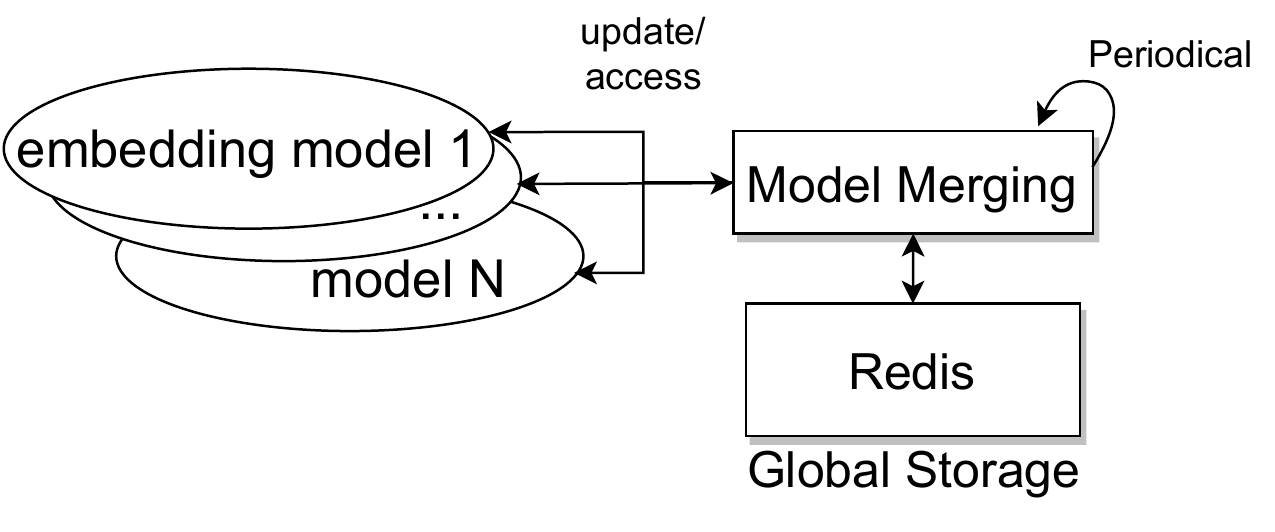}   
}    
\caption{Three model management strategies.}
\label{fig:modelmerge}
\end{figure*}
\subsection{Temporal Trend Detection}
\label{subsec:detection}
The conventional way of training word2vec assumes that an entire dataset is trained at once and does not consider the temporal attribute. 
It hence overlooks important information, such as temporally skewed opinions, which we call \emph{temporal semantic aggregation}. 
For example, there may be more positive texts (e.g. live tweets) timely aggregated because of an unbelievable shot in a football match. In addition, in product-level applications, continuously capturing trend can also help one keep an eye on real-time sentiment trends (such as latest reviews of a product).
To capture such important information, we leveraged a \emph{temporal trend detection} mechanism that helps our learning algorithm achieve higher prediction accuracy.

We illustrate the whole process of trend detection with an example of two trend detection windows ($TDWs$), shown in Figure~\ref{fig:trend}.
In each iteration, we identify the polarity of the current window of recent inputs (i.e., 100 tuples in this example). 
The trend detection results are used to feedback the classification of the next window of inputs by tuning the calculation of cosine similarity. 
For example, when the current window has a tendency to accumulate higher positive polarity, the weighted coefficient ($WC$) for calculating the positive polarity will become larger in the next window.
Both the trend detection window and the weighted coefficient adjustment step size are tunable, and we will evaluate the effect of TTD in our experiments (Section~\ref{subsec:system_sensitivity}).

\subsection{Periodical Model Merging}
\label{subsec:model_management}
A major limitation of modern DSPSs is the lack of support for \emph{shared mutable state}~\cite{tstream}, i.e., a shared memory that multiple application threads can access and update concurrently.
Shared state is crucial for \system as parallel training requires multiple threads to access and update the trained embedding model. 
This is not supported by Flink by default.
Commonly, one will leverage a key value store (KVS) to maintain such shared mutable states for stream processing. 
Typical examples of such KVS systems include
Redis~\cite{redis} and RocksDB~\cite{rocksdb}.
As our experimental results will demonstrate, naively relying on a KVS to manage the access and update of a trained classifier leads to a huge run-time overhead. 

We propose three alternative model management strategies in \system, as depicted in Figure~\ref{fig:modelmerge}.
1) \emph{local model:} 
we can maintain the model locally at each thread. This promises the best system performance as it does not involve costly Redis access. However, it raises the problem of divergence among learned models at each thread, and it can lead to lower prediction accuracy. Deploying the system in a distributed environment with multiple nodes would further exacerbate the problem.
2) \emph{global model:} 
we can share the learned model of word2vec among threads. This ensures a global model in the system that promises higher prediction accuracy. However, when the input stream is processed in a parallel fashion, different processing threads may subsequently access the same model concurrently, which can lead to significant access contention overheads. 
And 3) \emph{hybrid model:} 
we can periodically merge models among threads using a customizable frequency that allows users to trade-off accuracy for performance on demand. In particular, we compute the $mean$ of two vectors as the new vector representation of the corresponding word. Other pooling mechanisms~\cite{bommasani-etal-2020-interpreting} may be also applied such as $min$ and $max$. 

\compact
\section{Evaluation}
\label{sec:experiment}
In this section, we show that \system manages to achieve much better prediction accuracy, higher throughput, and lower processing latency compared to the state-of-the-art through a detailed experimental evaluation.

\subsection{Experimental Methodology}
In this section, we first introduce the benchmark workloads and evaluation metrics, followed by the discussion of several alternative approaches that we implemented in the same system to validate the efficiency of our proposal.

\subsubsection{Benchmark Workloads}
We evaluate \system on two well-known large-scale datasets~\cite{go2009twitter,zhangCharacterlevelConvolutionalNetworks2015} that are labelled with specific classifying rules. 
They cover different use cases and have distinct workload features. Sentiment140 contains mostly short texts, while Yelp Review contains relatively long texts. Note that, the original labels of the two datasets are only used to evaluate the prediction accuracy, and are not used for training in \system. 
\begin{myenumerate}
\item  
The \textbf{Sentiment140} dataset~\cite{go2009twitter} is extracted through the Twitter API from Twitter. 
It is labelled as positive if the tweet contains emoticons like ":)" or negative if tweet contains emoticons like ":(". The dataset contains 1.6 million random-topic tweets, and half are labelled as positive.
\item  
The \textbf{Yelp Review Polarity} dataset~\cite{zhangCharacterlevelConvolutionalNetworks2015} was extracted from the Yelp Dataset Challenge in 2015. Yelp review polarity consists of 560,000 training samples and 38,000 testing samples. Half are labelled as positive by considering reviews if they have stars 4 and 3; negative if they have stars 1 or 2. 
We combine training data (with labels removed) with test data as the input data stream in our experiment. Unless otherwise stated, we evaluate the \system on Yelp Review Polarity dataset in Section ~\ref{subsec:overall}, ~\ref{subsec:system_sensitivity} and ~\ref{subsec:scalability}.
\end{myenumerate}

    

\subsubsection{Evaluation Metrics} 
We evaluate the system with three performance metrics.
\begin{myitemize}
    \item \textbf{Throughput.} 
    Achieving high throughput is a must to keep up with large volume data streams. For example, when some major events happen, the opinions on social media may suddenly explode. 
    We measure the maximum number of input tuples per second that the system can sustain as the throughput.
    \item \textbf{Latency.} 
    We measure the 95\% latency as the elapsed time from when the input tuple arrives to when the corresponding classification result is produced. It is an important indicator to denote the system's responsiveness.
    \item \textbf{Accuracy.} 
    We define prediction accuracy as the proportion of correct predictions (summation of \emph{true positives} and \emph{true negatives}) over the total number of tuples processed. To evaluate the prediction accuracy for a workload with class imbalance, we also use the \emph{F1-score}, which is a harmonic mean of the precision and recall.
\end{myitemize}

\subsubsection{Alternative Approaches} 
To validate the superiority of our proposal, we implement and evaluate alternative unsupervised solutions including lexicon-based, probabilistic model-based (i.e., LDA), and clustering-based ones inside the same system.


\textbf{Lexicon.} 
Lexicon-based sentiment classification relies heavily on the quality of lexicons~\cite{2011Lexicon}. 
The core idea of this approach is to classify the polarities through scoring, by looking up the words in the lexicon and obtaining the corresponding scores. 
However, there is no lexicon that can cover all words because a lexicon is costly to build and requires intensive manual maintenance if it needs to be updated. 
Text-Blob~\cite{loria2018textblob}, VADER~\cite{hutto2014vader} and SentiWornet~\cite{baccianella2010sentiwordnet} are three well-known lexicons for unsupervised sentiment classification. 
Among them, the Vader lexicon contains only 7,500 features, while Sentiwordnet consists of more than 100,000 words and has a better ability to cover large-scale data than Vader and other lexicons. Therefore, we have implemented and evaluated Sentiwordnet in our experiments.

\textbf{LDA.} 
Latent dirichlet allocation (LDA)~\cite{blei2003latent} is widely used
for processing large-scale corpora. LDA determines the topic of each document in the document set in the form of a probability distribution. 
Based on LDA, Lin et al.~\cite{lin2009joint} proposed a Joint Sentiment Topic (JST) model that can infer sentiment by adding an additional sentiment layer between the document and topic layers. 
In conjunction with prior information from lexicons, JST has shown great performance on the IMDB movie reviews dataset (document-level) with an overall accuracy of 82\%, close to the performance of supervised approaches.
In our experiments, we have implemented and evaluated the performance of JST~\cite{JST}.


\textbf{Clustering.} 
Data clustering aims at grouping data such that objects in the same group are similar to each other and those in different groups are dissimilar according to the properties of the instances. 
In the literature, some have used clustering for unsupervised sentiment classification~\cite{li2014sentiment,Zimmermann:NeuroComp14}.
We have implemented the clustering-based approach based on streaming K-means, and set the $K$ parameter to 2 (i.e., either positive or negative) as another baseline method for sentiment classification in our evaluation. 



\subsubsection{Experimental Setup}
We have deployed our experiments on a prototyping cluster with four nodes.
Each node is equipped with 3.60GHz Intel(R) Xeon(R) CPU E5-1650 with 12 cores and 125GB RAM. The network bandwidth is 1Gb/s.
We have evaluated the system performance up to 48 working threads. 
We have used Flink v1.12.0 with Java 8 (Oracle JVM), Python 3.7 and Redis v6.0.6.
Finally, Table~\ref{tbl:reference} summarizes the hyperparameters of \system and their default values. 

\subsection{Overall Performance Comparison}
\label{subsec:overall}
In this section, we compare algorithms on processing benchmark workloads in terms of throughput, latency, and accuracy. 
Each algorithm is experimentally tuned to its best-performing configuration. 



\begin{figure}[t]
    \centering  
    \subfloat[Throughput]{
    \label{fig:thght}
    \includegraphics[width=0.4\textwidth]{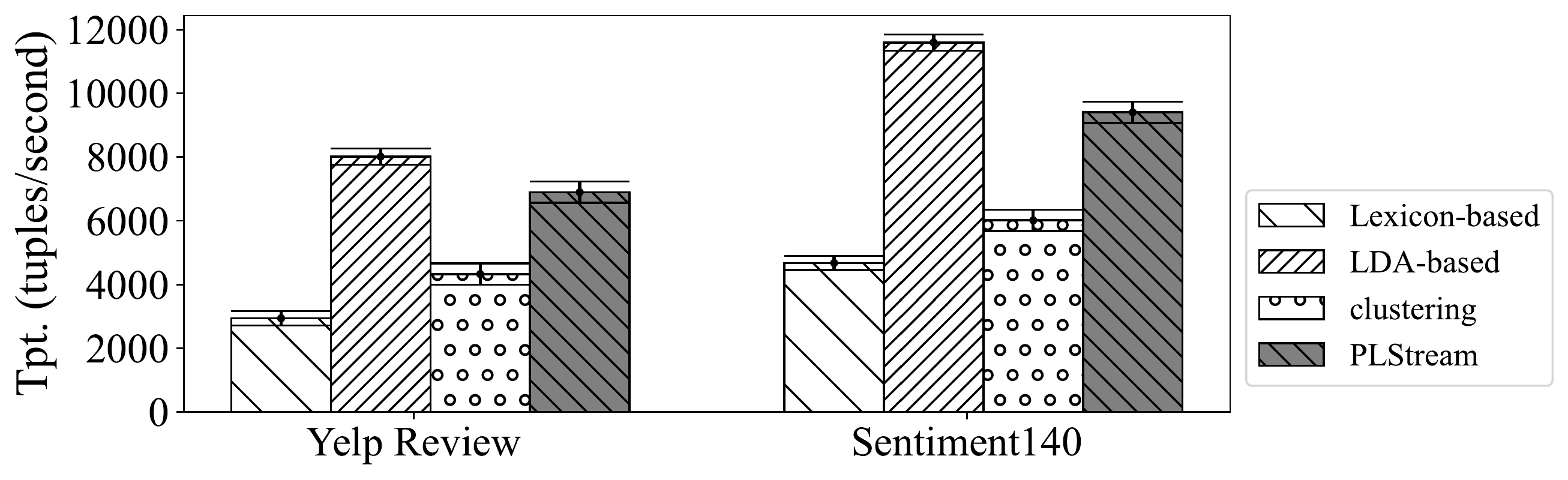}}
    \vspace{-9pt}
    \subfloat[$95^{th}$ Latency]{
    \label{fig:ltncy}
    \includegraphics[width=0.4\textwidth]{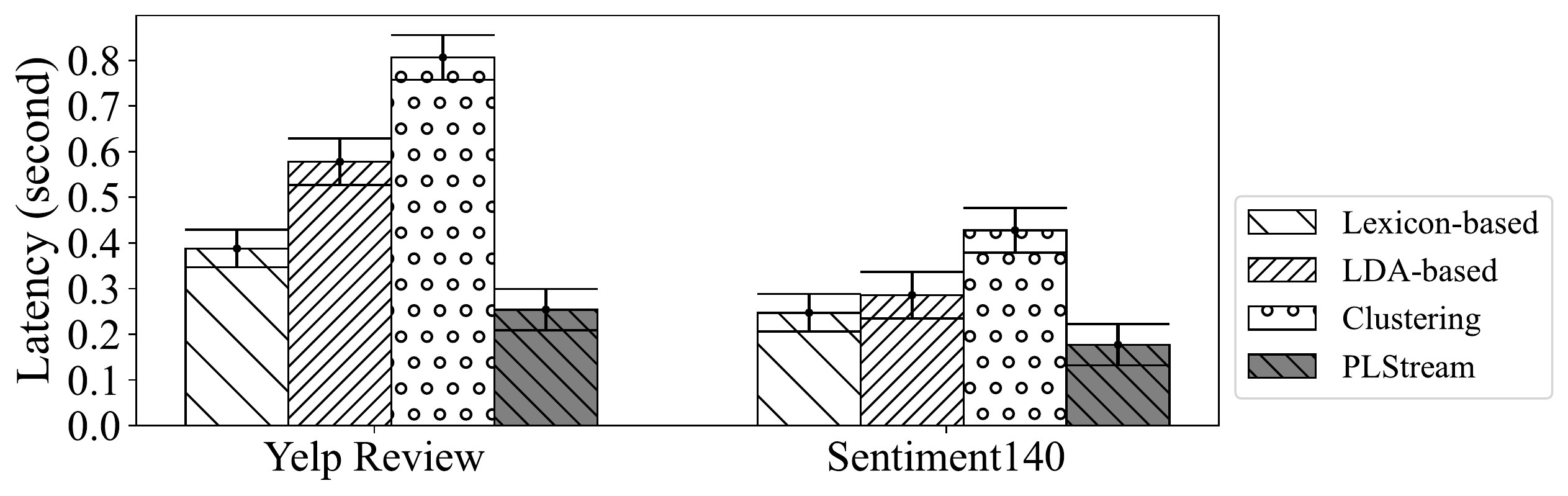}}
    \caption{Overall performance comparison.}
    \label{Fig.length}
\end{figure}

\textbf{Throughput and Latency.}
Figure~\ref{fig:thght} shows the throughput of alternative approaches on processing two datasets when the system is stable. The LDA-based method has the highest throughput because it only involves a simple inference stage in this case. Its performance may drop significantly when the prior information (i.e., the probability distribution) it requires needs to be re-calculated.
On the contrary, \system continuously updates its learned vector representation from new data, which introduces unavoidable overhead. 
Nevertheless, \system's throughtput is still promising, reaching 9877 tuples/second (Sentiment140) and 7182 tuples/second (Yelp Review) respectively.
Lexicon-based and clustering-based approaches achieve poor performance because the former involves costly lookups from the lexicon, while the latter involves a costly clustering computation process for every batch of input stream.
Figure~\ref{fig:ltncy} shows that \system achieves the lowest processing latency confirming the effectiveness of its fast incremental word embedding and low overhead similarity-based classification mechanisms. We will demonstrate the impact of each optimization techniques in detail later in Section~\ref{subsec:optimization}.

\begin{figure}[t]
    \centering  
    \subfloat[Sentiment140]{
    \label{Fig.sub.1}
    \includegraphics[width=0.4\textwidth]{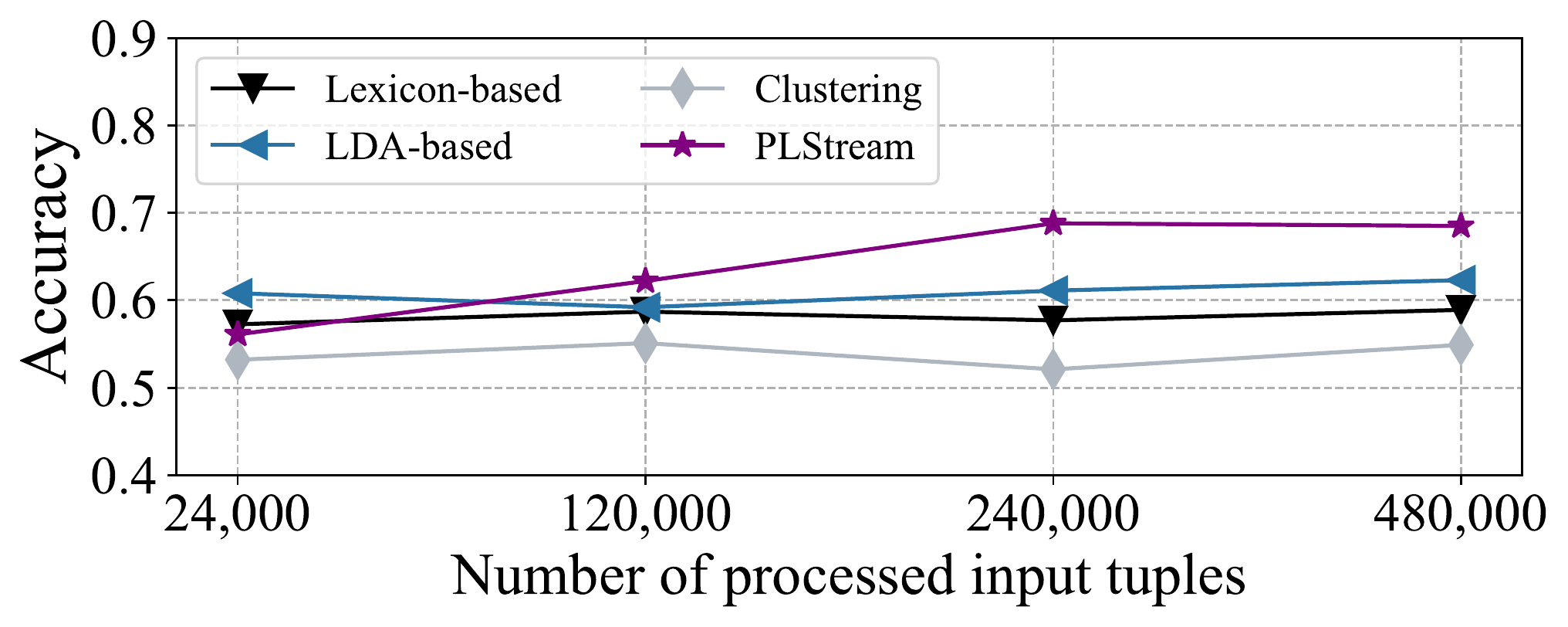}}
    \vspace{-9pt}
    \subfloat[Yelp Review]{
    \label{Fig.sub.2}
    \includegraphics[width=0.4\textwidth]{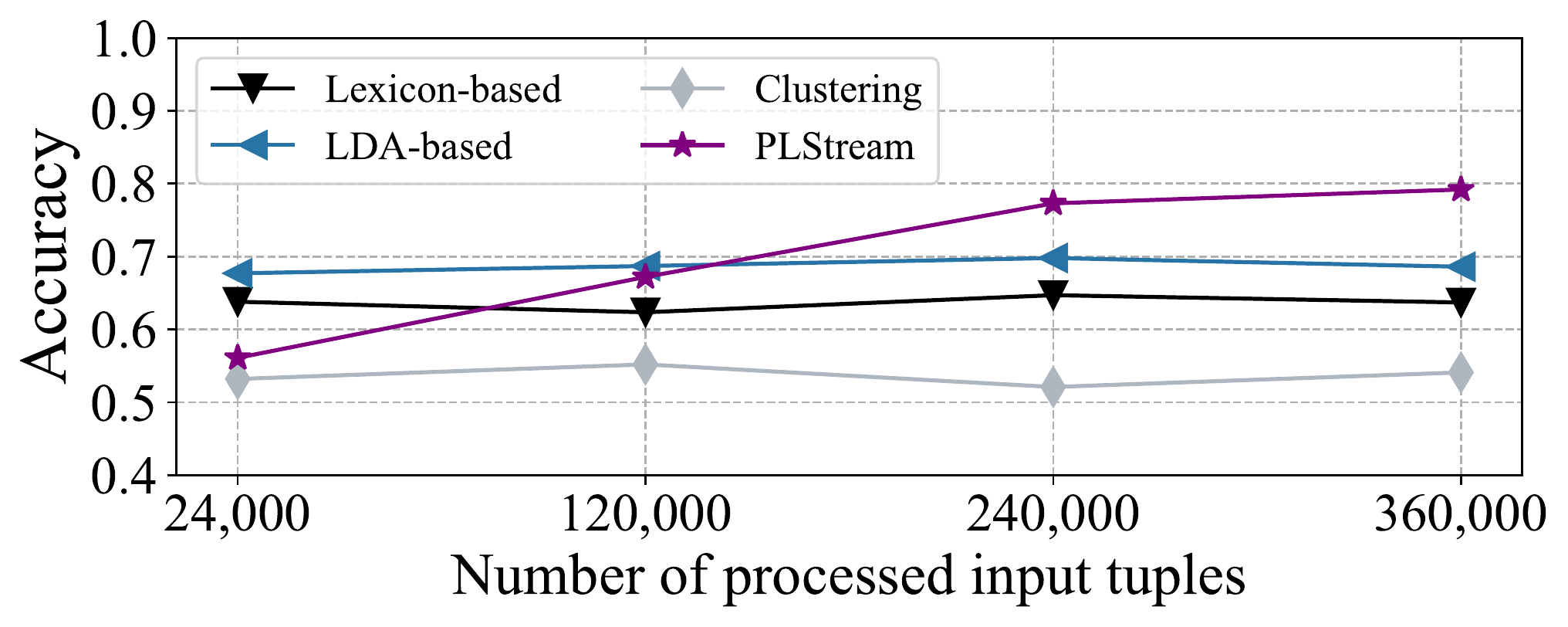}}
    \caption{Overall accuracy comparison.}
    \label{Fig.acc}
\end{figure}

\textbf{Accuracy.} 
Figure~\ref{Fig.acc} shows the evaluation of labelling accuracy. There are three key observations.
First, 
we can see that the accuracy of \system increases with more input tuples, since it can incrementally learn from new data, i.e., the quality of the vector representation improves gradually.
Starting from $360,000$ and $480,000$ input tuples, the accuracy begins to converge, reaching an average accuracy of 79.2\% and 68.8\% for Yelp Review and Sentiment140, respectively.
Second, 
the LDA-based method, which is sometimes treated as the representative unsupervised sentiment classification algorithm~\cite{liu2012survey}, achieves an accuracy of 68.5\% on average for handling the Yelp Review dataset. Its accuracy is even poorer when processing Sentiment140 (61.3\%), because it relies on a larger corpus for distribution computations of vocabulary and topics, while Sentiment140 contains mostly short texts.
Third, 
the clustering-based approach achieves the worst accuracy on both datasets. It achieves an average accuracy of 52.3\% and 51.9\% for Sentiment140 and Yelp Review, respectively. This is because the topics involved in the input stream are relatively random, and the text lacks important clustering features required for it to be effective.


\subsection{Impact of Optimizations}
\label{subsec:optimization}
A series of system optimization strategies further improve \system in terms of both accuracy and performance. 
In this section, we study the effectiveness of the key optimization techniques. 
Unless otherwise stated, we evaluate the labelling accuracy of \system on Yelp Review Polarity. We tune each optimization technique to its optimal configuration experimentally. 

\begin{figure}[]
    \centering  
    \subfloat{
    \label{Fig.sub.1}
    \includegraphics[width=0.5\textwidth]{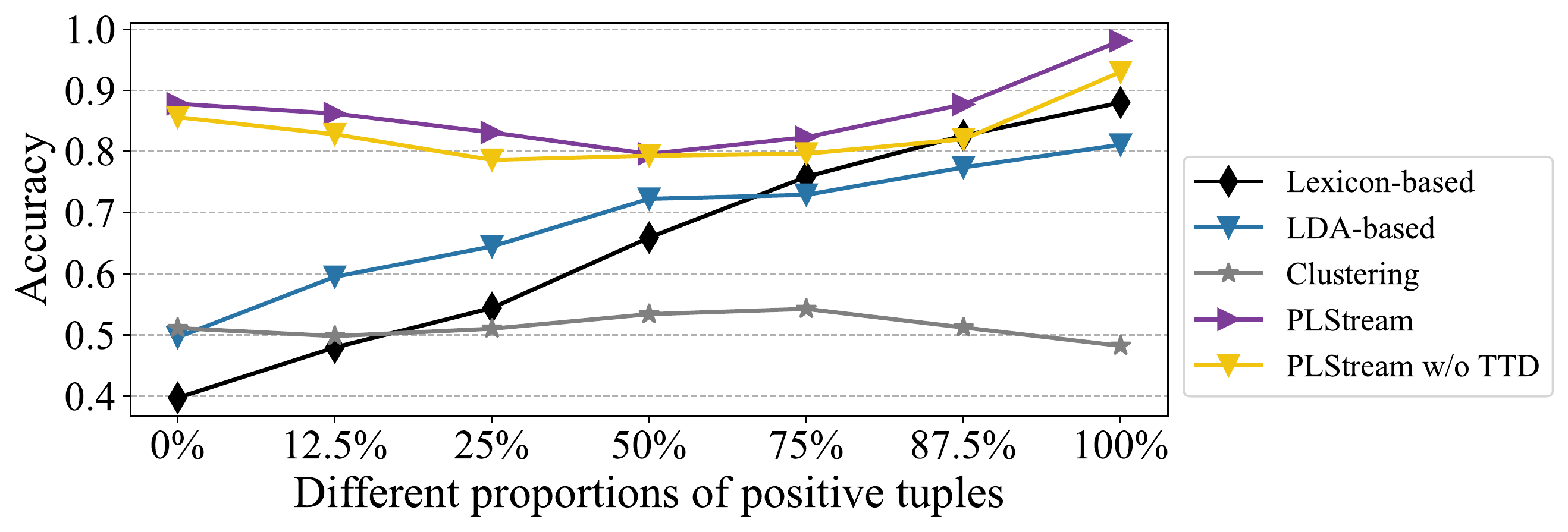}}
    \caption{Impact of temporal trend detection.}
    \label{Fig.distribution}
\end{figure}

\textbf{Impact of Temporal Trend Detection (TTD).}
We have generated datasets with more or less positive texts based on the Yelp Review dataset. The generated datasets have proportions of positive texts ranging from 0-100\%. 
We have evaluated the performance of various algorithms on these regenerated datasets in terms of accuracy and $F1-score$.  
It can be observed from Figure~\ref{Fig.distribution} that the accuracy of \system becomes higher with a more skewed polarity distribution. 
This is due to the \emph{temporal trend detection} technique employed by \system, which is able to react well to temporal semantic aggregation and improves the classification accuracy by up to $5.2\%$.
In contrast, the accuracy of lexicon- and LDA-based algorithms, is clearly lower for negative samples than for positive samples. 
This indicates that they misclassified a large number of samples, resulting in many \emph{false positives}. 
This is also reaffirmed by the $F1-score$ shown in Table~\ref{tab:f1}. 
Specifically, the Lexicon- and LDA-based algorithms have an $F1-score$ of less than 0.7 for the dataset with low positive samples, whereas \system outperforms all of them with an $F1-score$ greater than 0.8.

Our further investigation found that the trend detection window ($TDW$) has a significant impact on the effectiveness of the temporal trend detection technique. 
Specifically, the system performance with a smaller size $TDW$ (such as 10 or 20) becomes unstable. This is because prediction accuracy may be low in a small window, leading to low trend detection accuracy, which continuously affects the classification accuracy of subsequent windows. On the contrary, an oversized $TDW$ may miss the semantic aggregation. 
Automatically choosing the right size of $TDW$ is important and possible~\cite{2018Taking}, and we leave its tuning as a future work due to its considerable complexity.

\begin{table}[]
\caption{$F_1-score$ on different polarity distributions}
\label{tab:f1}
\setlength{\tabcolsep}{0.8mm}{
\small
\begin{tabular}{|c|c|c|c|c|c|c|c|}
\hline
\diagbox[innerwidth=2cm]{algo.}{pos. dist.}              & 0\%   & 12.5\% & 25\%   & 50\%  & 75\%  & 87.5\% & 100\% \\ \hline
Lexicon-based                                                      & 0.568 & 0.545  & 0.557  & 0.656 & 0.747 & 0.833  & 0.936 \\ \hline
LDA-based                                                          & 0.639 & 0.661  & 0.6667 & 0.731 & 0.734 & 0.793  & 0.876 \\ \hline
Clustering                                                         & 0.674 & 0.581  & 0.541  & 0.613 & 0.578 & 0.591  & 0.648 \\ \hline
\begin{tabular}[c]{@{}c@{}}PLStream \\ without TTD\end{tabular} & 0.922 & 0.853  & 0.798  & 0.803 & 0.812 & 0.837  & 0.975  \\ \hline
PLStream                                                        & 0.927 & 0.865  & 0.827  & 0.832 & 0.845 & 0.850  & 0.99  \\ \hline
\end{tabular}}
\end{table}

\begin{figure}[]
    \centering  
    \subfloat[Accuracy]{
    \label{Fig.sub.1}
    \includegraphics[width=0.4\textwidth]{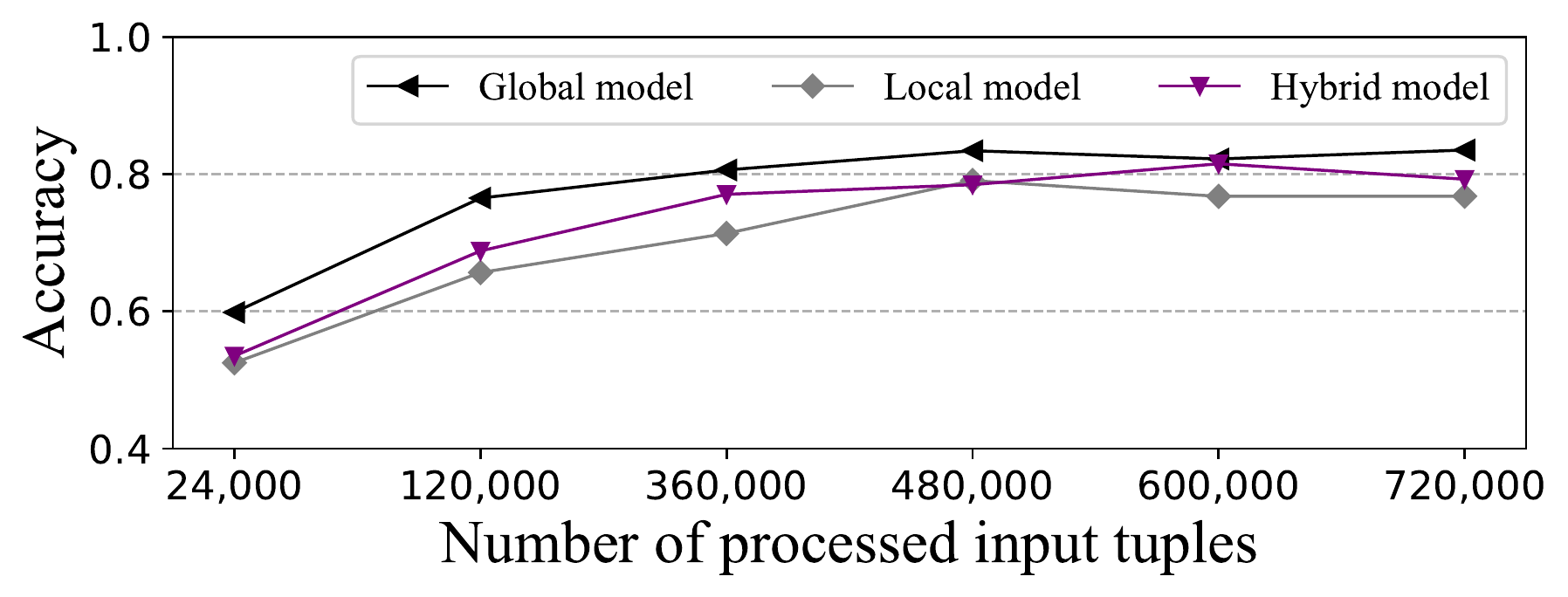}}
    \vspace{-9pt}
    \subfloat[Throughput]{
    \label{Fig.sub.2}
    \includegraphics[width=0.4\textwidth]{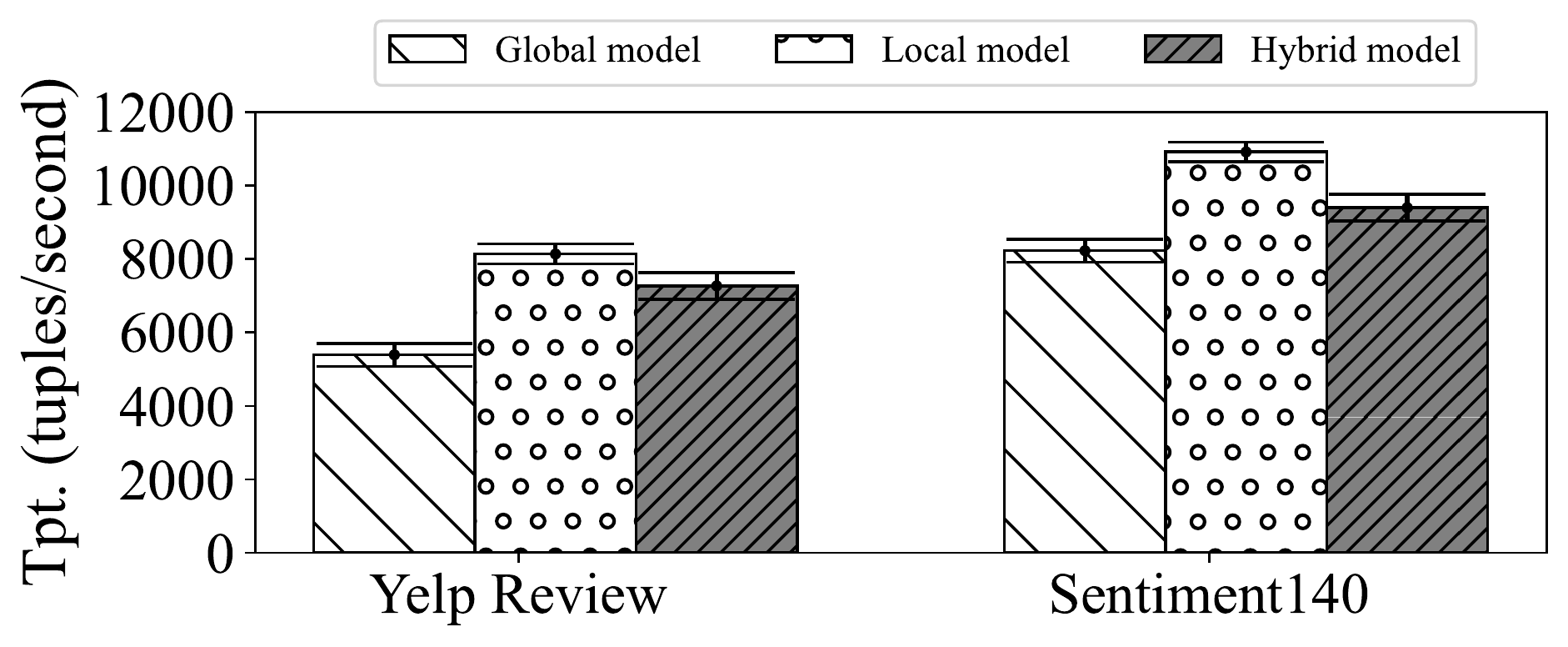}}
    \caption{Impact of model merging.}
    \label{Fig.modelmerge}
\end{figure}

\textbf{Impact of Model Merging.} 
Figure~\ref{Fig.modelmerge} clearly shows the trade-offs among three model management strategies in terms of accuracy and throughput. 
There are three observations.
First, the global model strategy achieves the highest prediction accuracy, as it generates the most comprehensive vector representations in the system. 
However, its throughput is the worst due to the contended global model updates to Redis.
Second, unlike the global model strategy, there is no communication between threads in the local model strategy, which leads to significantly higher throughput. However, it achieves the lowest classification accuracy and slowest convergence speed since the model on each thread can only learn from the data that is distributed to it. 
Third, as expected, the hybrid model achieves reliable accuracy and relatively good system performance at the same time. 
Our further evaluation has confirmed the clear trade-off of varying model merging frequency from 20 to 90 seconds. Higher frequency leads to better accuracy, but lower performance as frequent model merging can bring down the system performance. 
In the following experiments, we apply a hybrid model merging strategy with a merging frequency of 30 seconds.

\subsection{System Sensitivity Study}
\label{subsec:system_sensitivity}
In this section, we present the effectiveness of tuning various hyperparameters of \system in detail. 

\begin{figure}[]
\centering
    \subfloat[Accuracy]{%
        \includegraphics*[width=0.4\textwidth]{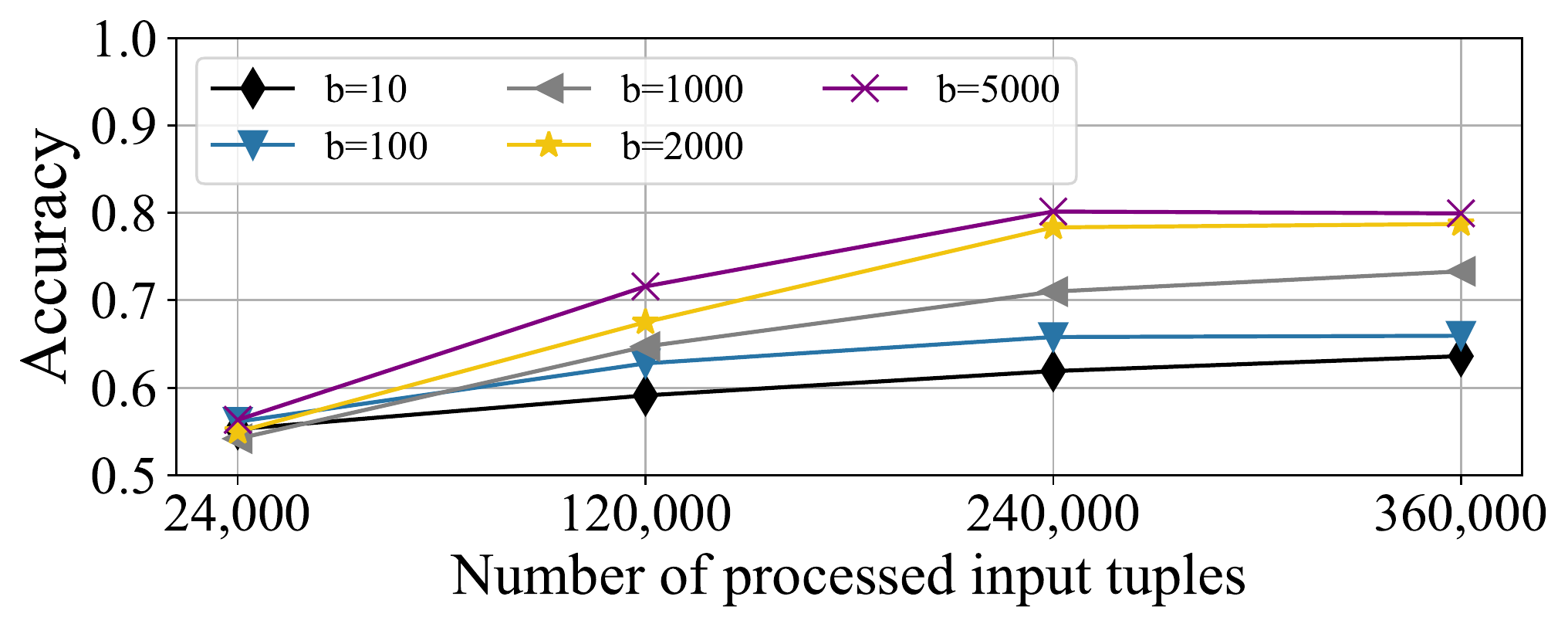}   
    }     
    \begin{center}
     \fbox{
     \includegraphics[width=0.85\columnwidth]
         {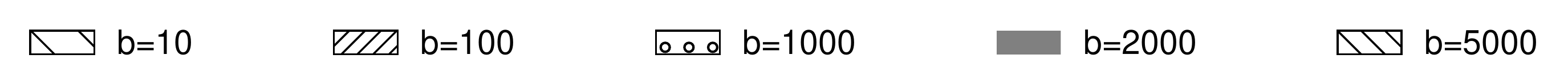}
     }
     \vspace{-4pt}
    \end{center}
    \subfloat[$95^{th}$ Latency]{%
        \includegraphics*[width=0.235\textwidth]{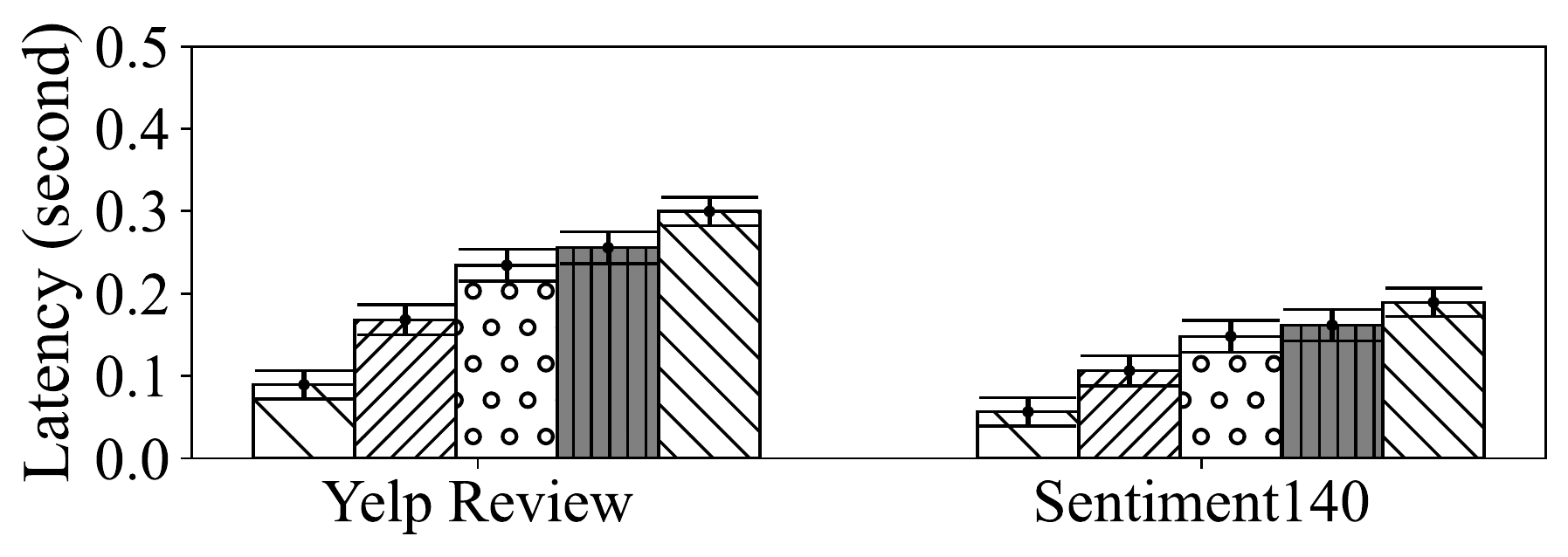}   
    }
    \subfloat[Throughput]{%
        \includegraphics*[width=0.22\textwidth]{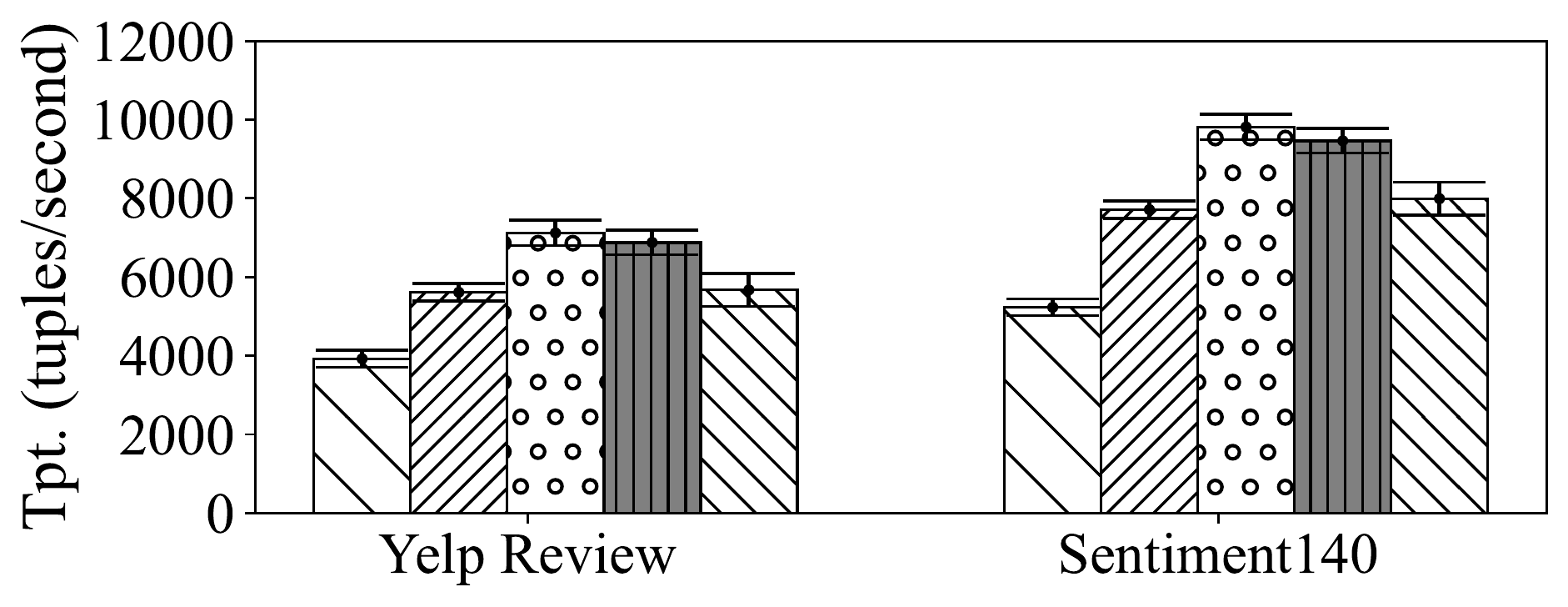}   
    }    
    \caption{Impact of adaptive batching ($b$).}
    \label{fig:collector_size}             
\end{figure}

\textbf{Impact of Adaptive Batching ($b$).}
A larger batch size allows \system to learn from more input tuples at a time. Hence, it achieves a faster convergence speed, lower latency, and higher throughput with amortized function call cost, as shown in Figure~\ref{fig:collector_size}. 
However, there is a clear decrease in system performance when $b$ exceeds 2000. 
The reason is that each thread in \system needs to wait for more tuples to start training, resulting in more system idle time. 

\begin{figure}
\centering
    \subfloat[Accuracy]{%
        \includegraphics*[width=0.4\textwidth]{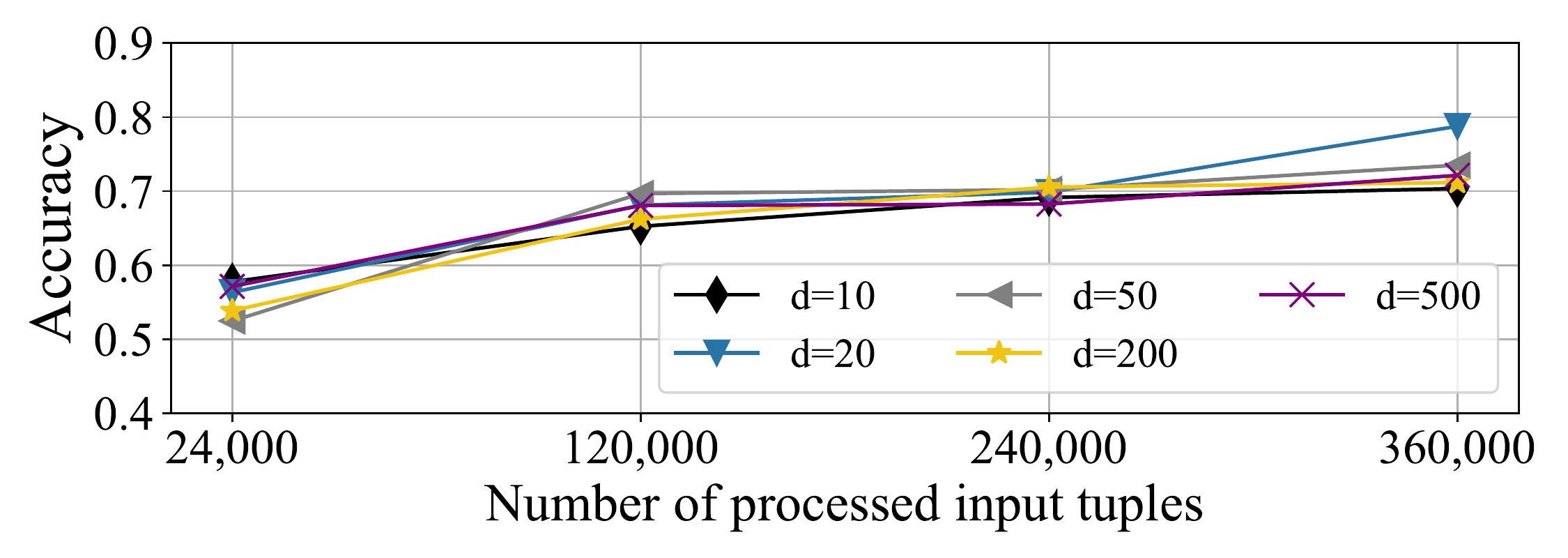}   
    }     
    \begin{center}
     \fbox{
     \includegraphics[width=0.85\columnwidth]
         {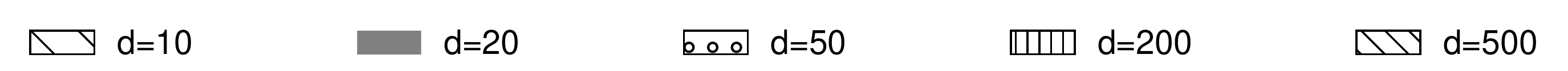}
     }
     \vspace{-5pt}
    \end{center}
    \subfloat[$95^{th}$ Latency]{%
        \includegraphics*[width=0.235\textwidth]{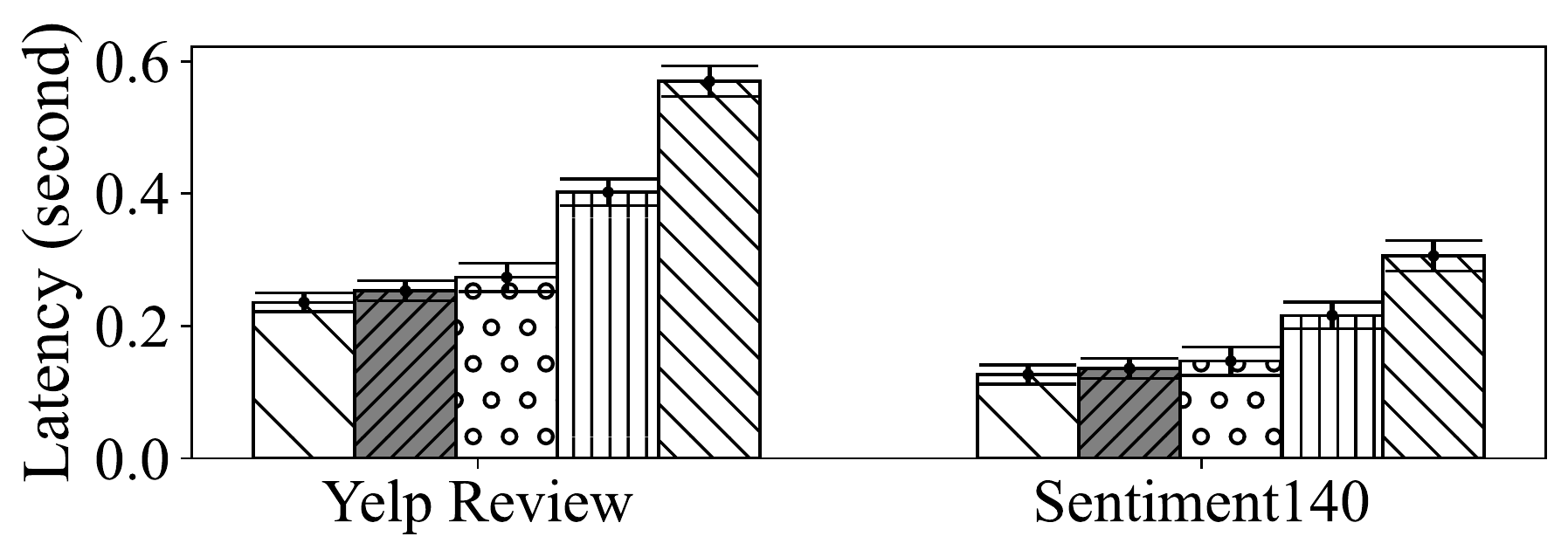}   
    }
    \subfloat[Throughput]{%
        \includegraphics*[width=0.235\textwidth]{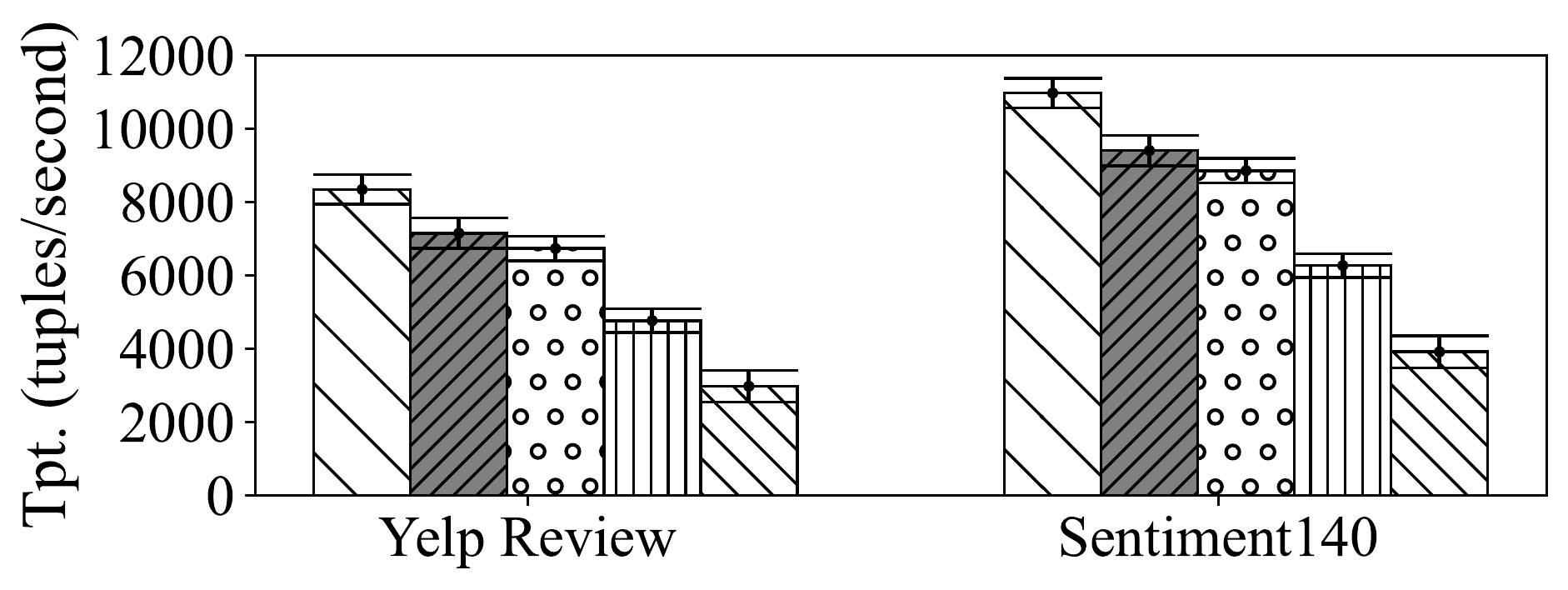}   
    }    
    \caption{Effect of varying vector dimension ($d$).}
    \label{fig:vectors}             
\end{figure}

\textbf{The Effect of Varying Vector Dimension ($d$).}
We may tune the number of features (i.e., vector dimension $d$) used in word2vec training. 
The common wisdom is that a large vector dimension ($\sim$500~\cite{patel-bhattacharyya-2017-towards,mikolov2013efficient}) helps to obtain a better quality vector representation.
Figure~\ref{fig:vectors} (a) illustrates how the classification accuracy changes by varying $d$ from 10 to 500. 
We can see that the converged accuracy (when the number of processed input tuples=30k) increases from $d=10$ to $d=20$. This matches the previous findings~\cite{patel-bhattacharyya-2017-towards}.
However, the classification accuracy does not increase and sometimes even drops once $d$ exceeds 20.
This is due to the well-known \emph{curse of dimensionality}~\cite{10.1007/11494669_93} problem raised during our similarity-based classification. In particular, when the dimensionality is high, data points (i.e., vector representations in our case) are mostly sparse and disimilar, which prevents reliable detection of data points of similar properties.
Figure~\ref{fig:vectors} (b) and (c) show that as the vector dimension becomes higher, \system achieves lower throughput and higher processing latency. 
This is due to the increasing cost spend in matrix multiplications during incremental word vectorization, as well as the costly similarity-based classification.

\subcompact
\subsection{Solution Scalability}
\label{subsec:scalability}
In this section, we evaluate the scalability of \system with respect to hardware resources and workload complexity.

\begin{figure}
\centering
    \subfloat[Accuracy]{%
        \includegraphics*[width=0.4\textwidth]{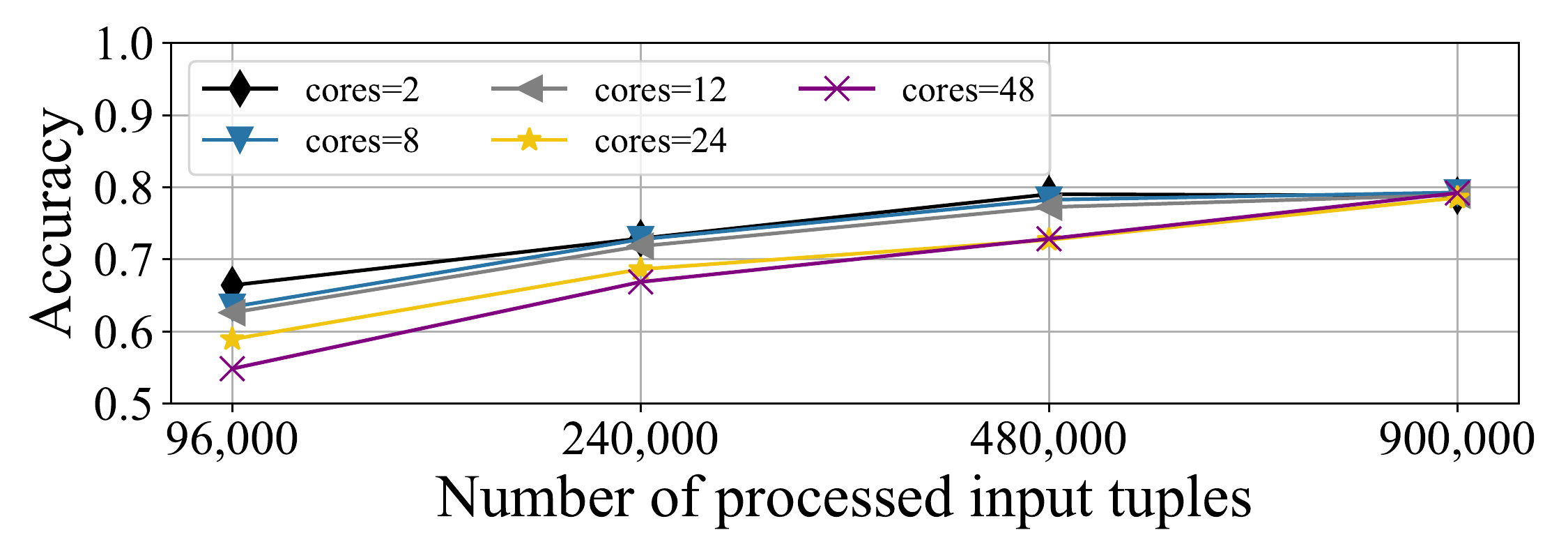}   
    }     
    \begin{center}
    \end{center}
    \subfloat[$95^{th}$ Latency]{%
        \includegraphics*[width=0.235\textwidth]{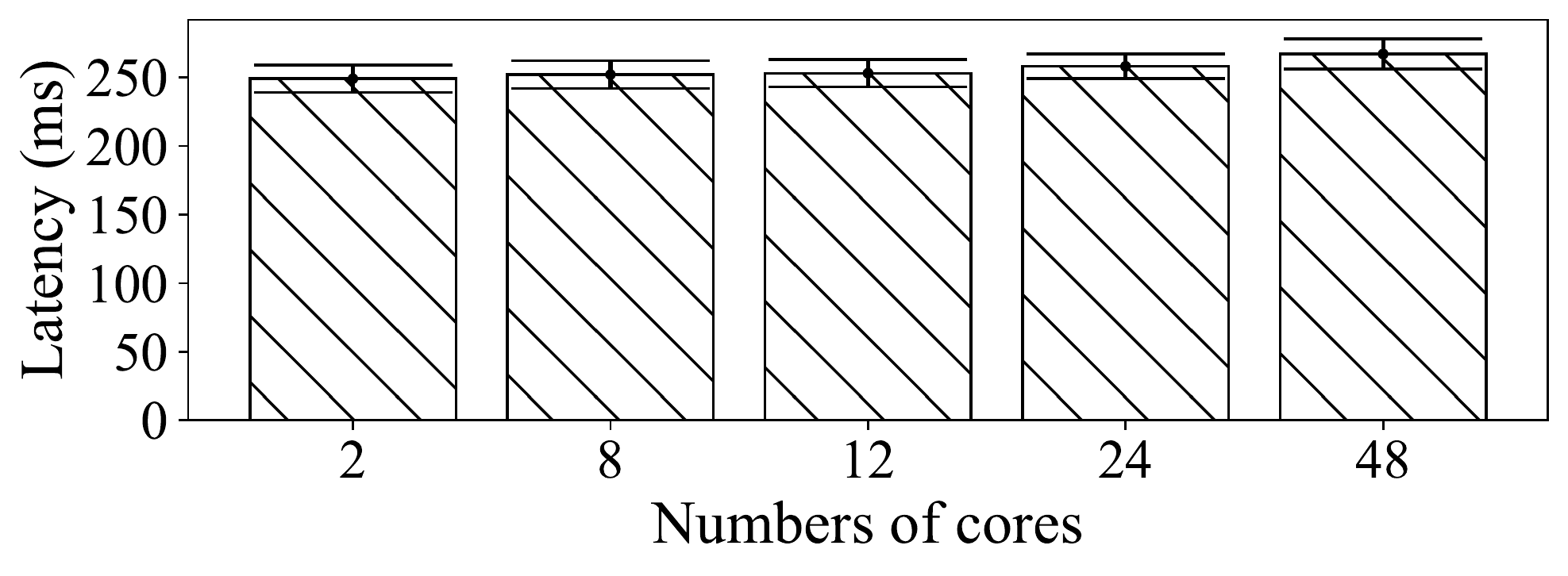}   
    }
    \subfloat[Throughput]{%
        \includegraphics*[width=0.235\textwidth]{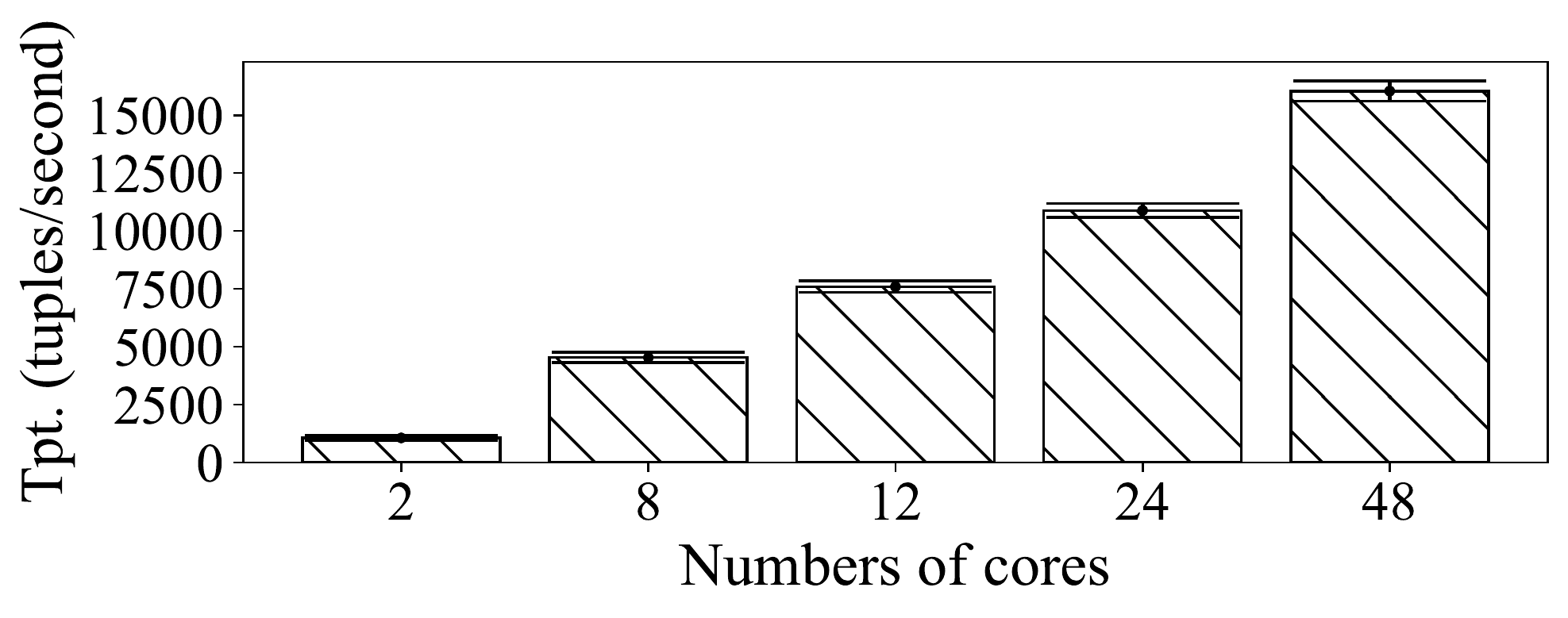}   
    }    
    \caption{Effect of varying number of CPU cores.}
    \label{fig:scale}             
\end{figure}

\textbf{Hardware Resource Scalability.}
Our next experiment shows that \system scales effectively , when we increase the number of CPU cores. 
In this experiment, \system is configured with hybrid model strategy and LRU-based model pruning technique.
Figure~\ref{fig:scale} (a) shows that as the number of threads increases, \system needs to process more tuples to achieve accuracy convergence. This is expected, as each thread learns less with increasing number of threads. 
In contrast, \system's processing and incremental learning at different threads are mostly independent with little contention among others. 
In particular, we can see from Figure~\ref{fig:scale}(b) that the processing latency of \system does not vary with the different number of cores in one node. The processing latency increases marginally when \system is deployed on multiple nodes due to the additional network communication overhead. 
Figure~\ref{fig:scale} (c) further shows that the throughput of \system improves significantly when increasing the number of cores from 2 to 12 cores (one node), up to 48 cores (four nodes), reaffirming its good scalability. 
Nevertheless, further exploration is required to improve the system towards linearly scale-up.

\begin{figure}[]
    \centering
    \includegraphics[width=0.4\textwidth]{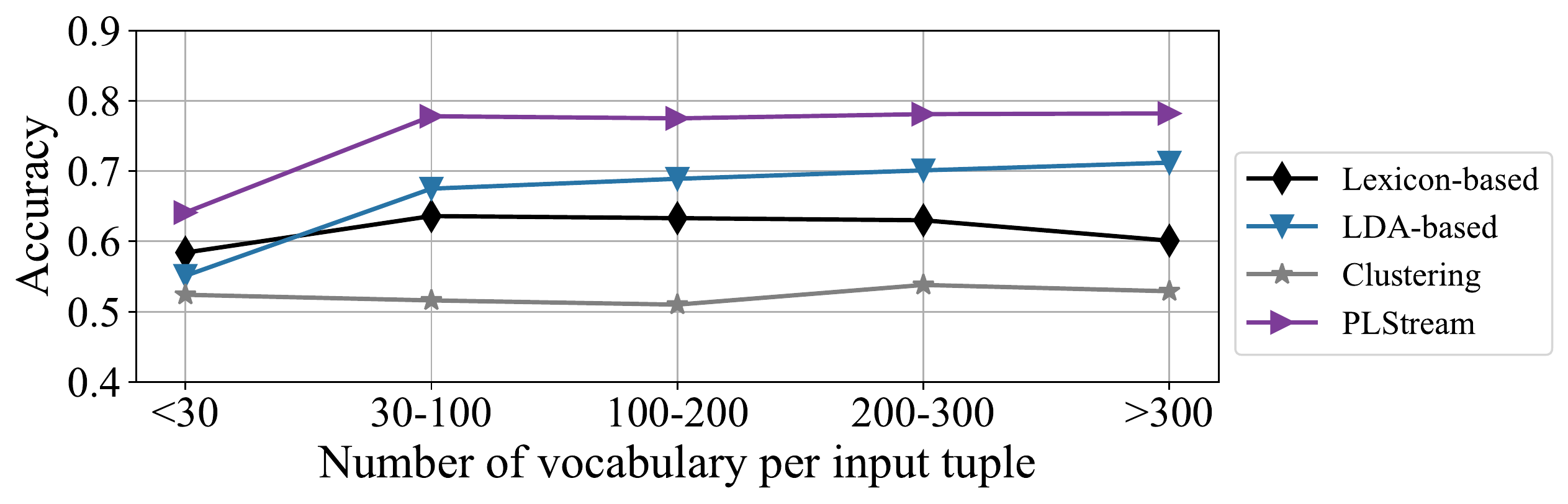}
    \caption{Accuracy on different text length.}
    \label{Fig.length_acc}
\end{figure}

\textbf{Workload Complexity Scalability.}
Data streams such as reviews have longer text, whereas tweets, for example, are dominated by short text. We now examine our annontating effect in terms of text length.
We selectively pick texts of varying length from the two benchmark workloads to regenerate the input stream for this experiment.
Figure~\ref{Fig.length_acc} shows the converged accuracy of each algorithm on the data set of different text lengths. We have repeated each experiment three times and report the average result. We have four major observations.
First, it can be observed that the accuracy of \system always outperforms other algorithms. Interestingly, in the dataset with a text length of less than 30, the accuracy of all algorithms is poorer. 
This is because, unlike paragraphs or documents, short texts are more ambiguous: they do not contain enough contextual information, which poses a great challenge for classification~\cite{Chen_Hu_Liu_Xiao_Jiang_2019}. 
Second, the accuracy of the LDA-based algorithm improves with increased text length.. This is expected, as the LDA-based approach is designed for document-level sentiment analysis~\cite{lin2009joint}.
Third, the accuracy of the lexicon-based algorithm also increases initially due to the increased contextual information in the input workloads. However, its accuracy decreases when the text length exceeds 300. This is due to the increasing noise interfering with its scoring process. For example, one input tuple may contain multiple opinions. 
Fourth, in contrast, varying text length shows no impact on the clustering-based approach. This is again due to the lack of important clustering features in our benchmark workloads.


\compact
\section{Conclusion}
\label{sec:conclusion}
We have presented \system, a new framework designed for annotating fast ongoing \emph{unlabelled} data streams at \emph{scale} on modern parallel machines. 
\system encompasses several algorithmic and system architectural innovations, that reduce workloads, enhance learned embedding models incrementally, and control model management overhead. Based on extensive experiments with two real-world workloads, we demonstrate \system's high labelling accuracy, system throughput, and low processing latency. 
In practice, users can further use a confidence threshold to control the labelling of the vast majority of the data with high confident by \system and leave the rest to human experts. Labelled datasets can be immediately consumed by subsequent supervised learning tasks and reduce the development cost of sentiment analysis services. 
In the future, we plan to further exploiting ``auto-ML'' techniques to identify the optimal value of the various tuning knobs of \system.

\compact



\bibliographystyle{ACM-Reference-Format}
\bibliography{mylib}

\end{document}